\documentclass[reprint,nofootinbib,nobibnotes,amsmath,amssymb,aps,prb,twocolumn,floatfix]{revtex4}
\usepackage{epsfig}
\usepackage[dvipsnames,usenames]{color}
\usepackage{color}
\usepackage{amsmath}
\usepackage{amsfonts} 
\usepackage{comment}
\usepackage{array}
\usepackage{multirow}
\usepackage{tabularx}
\usepackage{float}
\usepackage{hyperref}
\usepackage{float}

\newcommand{\bra}[1]{|#1\rangle}
\newcommand{\ket}[1]{\langle#1|}
\newcommand{\br}{{\bf r}}
\newcommand{\brp}{{\bf r}^\prime}

\newcommand{\pdag}{{\phantom{\dagger}}}
\emergencystretch=\maxdimen
\begin{document}
\title{Polaron and bipolaron tendencies in a semiclassical model for hole-doped bismuthates}

\author{Mi Jiang,$^{1,2}$  George A. Sawatzky,$^{1,2}$ Mona Berciu,$^{1,2}$ and 
Steven Johnston$^{3,4,*}$} 
\affiliation{$^1$Department of Physics and Astronomy, University of
British Columbia, Vancouver B.C. V6T 1Z1, Canada} 
\affiliation{$^2$Stewart Blusson Quantum Matter Institute, University of
British Columbia, Vancouver B.C. V6T 1Z4, Canada}
\affiliation{$^3$Department of Physics and Astronomy, The University of Tennessee, Knoxville, Tennessee 37996, USA} 
\affiliation{$^4$Joint Institute for Advanced Materials at The University of Tennessee, Knoxville, Tennessee 37996, USA}

\email[Correspondence: ]{sjohn145@utk.edu}

\begin{abstract}
Bismuth perovskites ABiO$_3$ (A = Sr, Ba) host a variety of peculiar phenomena including bond-disproportionated insulating phases and high-temperature superconductivity upon hole doping. While the mechanisms underlying these phenomena are still debated, off-diagonal electron-phonon ($e$-ph) coupling originating from the modulation of the orbital overlaps has emerged as a promising candidate. Here, we employ classical Monte Carlo simulations to study a semiclassical three-orbital model with off-diagonal $e$-ph interactions. We demonstrate the existence of a (bi)polaron correlations that persists in the model at high temperatures and for hole doping away from the bond-disproportionated insulating phase. Using a spatiotemporal regression analysis between various local quantities and the lattice degrees of freedom, we also identify the similarity between heating- and doping-induced melting of a bond-disproportionated insulator at a microscopic level. Our results imply that (bi)polaron physics can be a unifying concept that helps us understand the rich bismuth perovskite phase diagram. 
\end{abstract}

\maketitle

\section{Introduction}
Bond-disproportionated or ``dimerized'' insulating phases have been found in several families of quantum materials including the bismuth perovskites ABiO$_3$ (A = Sr, Ba)~\cite{Sleight1976, Cox1979,  Sleight2015, Foyevtsova2015} and the rare-earth nickelates RNiO$_3$~\cite{Medarde, George2000, Millis2012, Millis2013, Johnston2014}. In this phase, the oxygen octahedra surrounding the perovskite cations exhibit a long-range alternating in and out ``breathing'' distortion along the three cubic crystallographic directions. In the case of barium bismuthate, subsequent doping with Pb or K suppresses the insulating phase, relaxes the bond-disproportionated structure, and ultimately results in  high-temperature (high-T$_\mathrm{c}$) superconductivity. The rare-earth nickelates, on the other hand, cannot be hole doped. Still, they do undergo an insulator-to-metal transition at high temperature, with a similar relaxation of the bond-disproportionated structure.  

The literature invokes two main mechanisms -- charge disproportionation and bond disproportionation --  as the potential driving force behind these transitions. Here, we focus on the bismuthates, but these concepts also apply to the rare-earth nickelates with some caveats. 

The charge disproportionation scenario is built on the idea that the nominal Bi 4+ valence state of the valence-skipped Bi ions is not energetically favorable. Instead, charge transfer occurs that produces alternating Bi$^{3+}$ and Bi$^{5+}$ ions on interpenetrating sublattices.\footnote{In the case the rare-earth nickelates, charge disproportionation is facilitated by the Ni $3d$ Hund's coupling, which offsets the substantial Coulombic cost of the charge density modulation.~\cite{Mazin2007}} Short Bi$^{5+}$-O and long Bi$^{3+}$-O bonds form as a consequence, creating a dimerized structure. Once it forms, the wouldbe Bi $6s^1$ electrons then preferentially occupying one sublattice, giving rise to an insulating CDW state~\cite{Sleight1976, Cox1979, Rice1981, Varma1988,  Sleight2015, Yanagisawa2007, Franchini2009}. This scenario is, however, at odds with the the highly covalent nature of the Bi-O bond  and any local Coulomb repulsion active in the system~\cite{Mattheiss1983, Harrison2006}. Various spectroscopic techniques also find evidence for small charge differences between the two inequivalent bismuth ions~\cite{Hair1973,Orchard1977,Wertheim1982}, which speaks against this scenario.

Bond disproportionation is an alternative to charge disproportionation. This scenario derives from the notion that the bismuthates belong to the so-called ``negative charge transfer'' family of materials~\cite{Khomski1997, Ignatov1996, Menushenkov2000, Plumb2016}, where holes self-dope from the cation to the ligand oxygen atoms due to a low or negative value of the charge transfer energy $\Delta$.  In the bismuthates, this process converts the energetically unfavorable Bi $4+$ ion into a Bi$^{3+} \underline{L}$ configuration, where $\underline{L}$ denotes a hole on the ligand oxygen atoms. The energy of this configuration can be lowered further by coupling to a coherent breathing distortion of the lattice, thus forming a bond-disproportionated structure~\cite{Ignatov2000, Millis2012, Millis2013, Johnston2014}. In this scenario, all of the oxygen atoms are identical in terms of their orbital occupations, and pairs of holes occupy a molecular orbital of $A_{1g}$ symmetry. The resulting nominal valence of the Bi ions are all close to being $6s^2$ (i.e. $2 \rm Bi^{3+} \underline{L} \rightarrow \rm Bi^{3+} \underline{L}^2 + \rm Bi^{3+}$)~\cite{Ignatov2000}. The relevance 
of this scenario for the bismuthates has also been supported by more recent density-functional theory (DFT) ~\cite{Foyevtsova2015, Plumb2016, Shadi2017, Arash2018a} and model quantum Monte Carlo (QMC)~\cite{Steve2020} calculations. 

Charge and bond disproportionation are not mutually exclusive, of course, and it is possible that both a charge ordering on the Bi ions and lattice distortions contribute to the formation of the charge density wave gap in bismuth perovskites. In fact, these scenarios can be viewed as opposite ends of a range of possibilities controlled by the charge transfer energy and the size of the Bi-O overlap integral~\cite{Arash2018b}. For large positive $\Delta$, the holes would localize on the Bi sites, and the charge disproportionation mechanism would dominate. In this limit, one can describe the low-energy physics of the system using an effective  single-band model with a negative-$U$~\cite{Rice1981,Varma1988}. Conversely, as $\Delta$ decreases, the Bi holes will delocalize and transfer entirely to the oxygen orbitals once $\Delta$ becomes small enough. The charge disproportionation scenario dominates in this limit, and retaining the oxygen degrees of freedom becomes essential to describe the physics of the system. 

Determining where the nickelates and bismuthates lay on this spectrum has implications for our understanding of their metallic phase, and potentially superconductivity in the bismuthates~\cite{Sleight1975, Johnson1988, Kometani1988, Sleight2015, Shiroka2020}. The absence of magnetism in the bismuthate phase diagram implies that a nonmagnetic pairing mechanism is at work in these materials. Indeed, many ideas have been advanced, ranging from effective negative-$U$ centers arising from charge disproportionation~\cite{Varma1988, Ramakrishnan1995} to strong electron-phonon ($e$-ph) coupling~\cite{Rice1981, Savrasov1998} (possibly enhanced by correlations~\cite{ZPYin2013, Louie2019}) to high-energy charge excitations~\cite{Espinosa1988}. 

Angle-resolved photoemission spectroscopy (ARPES) experiments on  bismuthate single crystals (films) have recently become available, and have provided some insights into these questions. The first ARPES study on the parent compound BaBiO$_3$ found that its band structure was well described by DFT within the local density approximation (LDA) and was fully consistent with the negative charge transfer/bond disproportionate view~\cite{Plumb2016}. A later study on Ba$_{1-x}$K$_x$BiO$_3$ (BKBO) with $x=0.49$ observed an increased bandwidth~\cite{DLFeng2018} consistent with the theoretical prediction~\cite{ZPYin2013} that the long-range Coulomb interaction increases the bandwidth and enhances the $e$-ph coupling to generate a high T$_\mathrm{c}$. Another ARPES study on BKBO films interpreted the normal state spectra in terms of nanoscale phase separation and a polaron-liquid-like metallic phase~\cite{Naamneh2018}. This latter result is consistent with  the experimental observations of local short-range Ni-O bond distortions in the metallic phase of the related rare-earth nickelates~\cite{Li2016, Shamblin2018}.

This progress has motivated an extensive theoretical investigations aimed at extracting the relevant minimal models~\cite{Rice1981, Varma1988, Yanagisawa2007,  Franchini2009, Arash2018a,  Shamblin2018, Steve2020}. A recent DFT investigation supporting the bond disproportionation  scenario~\cite{Foyevtsova2015} has also motivated the derivation of several tight-binding models to describe the low-energy electronic excitations of the bismuthates and  their possible simplifications as a function of structural distortions in two and three dimensions~\cite{Arash2018a}. The most significant ingredient in these simplified models is the molecular orbital states straddling the Fermi energy. In the bismuthates, these molecular orbitals are of $A_{1g}$ symmetry relative to the central Bi ion.\footnote{The molecular orbital viewpoint seems to be quite general in high oxidation state oxides, e.g. nickelates with a linear combination of O-2p atomic orbitals of $E_g$ symmetry\cite{Millis2013, Johnston2014}, cuprates with the formation of the Zhang-Rice singlets~\cite{ZhangRice}.} 
The formation of these molecular orbitals provides a natural framework for the bond disproportion scenario and polaron physics. Here, molecular orbitals with a bonding symmetry will hybridize with the valence orbitals on central cation sites, and the subsequent breathing motion of the oxygen couples strongly to the carriers via the modulation of the cation-anion overlap integrals. In this framework, one expects that pairs of holes will be bound to local compressions of the oxygen octahedra forming small polarons~\cite{Shamblin2018, Steve2020}. The undoped bond-disproportionated insulating phase is then a frozen bipolaronic crystal state, which is melted by hole doping, eventually leading to a metallic bipolaron liquid characterized by fluctuating patches of local distortions. 

The microscopic $e$-ph coupling needed to capture the situation described is a generalization of the Su-Schrieffer-Heeger (SSH) interaction, where the motion of the oxygen atoms  
modulates the Bi-O overlap integrals.\footnote{Interactions of this type are also sometimes refereed to as ``Peierl's" couplings. For brevity, we will referee to them as SSH-like interactions, even though the original SSH model was derived for one-dimensional chains with hoping between identical orbitals.} This coupling mechanism leads to an $e$-ph interaction that is off-diagonal in orbital space, and is difficult to treat using exact nonperturbative methods on extended 2D and 3D lattices with finite carrier concentrations. Only recently have quantum Monte Carlo (QMC) calculations been possible for single- and multi-orbital models with SSH-like couplings in 2D~\cite{Steve2020,Scalettar2020}. Relevant to our discussion is the recent DQMC simulations of a 2D three-orbital SSH model, which examined the problem in the negative charge transfer limit~\cite{Steve2020}. There, the authors inferred a phase diagram qualitatively consistent with BKBO, with a dimerized insulating phase near half-filling and superconductivity appearing at larger hole concentrations. Importantly, they also found that the metallic phase was characterized by fluctuating patches of local distortions, consistent with a (bi)polaron-liquid-like state. That study was limited to a $4\times 4$ clusters, however, due to the long autocorrelation time associated with DQMC simulations of e-ph models~\cite{Li2019}, and the high computational costs of treating the SSH interaction. 

In this paper, we present a complementary study of a model for the bismuthates, similar to the one used in Ref.~\cite{Steve2020}. Here, however, we treat lattice degreess of freedom semiclassically and using a combined exact diagonalization and classical Monte Carlo approach. This method allows us to overcome the aforementioned system size limitations, and examine the phonons in the adiabatic regime. Using this approach, we study the polaron and bipolaron correlations in the model as a function of doping and temperature, in order to draw a complementary picture of the bond-disproportion mechanism in the bismuthates and other negative charge transfer oxides. 

This paper is organized as follows. Sec.~II describes our model and the combined exact diagonalization + classical Monte Carlo method used in this work. Sec.~\ref{Sec:Results} presents our results. First, Sec.~\ref{Sec:Insulating} establishes  
that the model has insulating properties at low temperature when there is on average one hole/Bi ion (i.e. at `` half-filling''). Next, Sec.~\ref{Sec:Results_Local} examines the formation of the bond-disproportionate state at and close to half-filling.  Sec.~\ref{Sec:Results_Bipolaron} discusses the evidence for bipolaron formation in the model as a function of doping and temperature, as seen from the perspective of several local quantities. This section also discusses our results in comparison to those obtained from the DQMC treatment of a similar model~\cite{Steve2020}. Secs.~\ref{Sec:Results_Histo} and \ref{Sec:Results_Histo2} present a histogram analysis of our Monte Carlo configurations. The purpose of this analysis is to provide a clearer picture of the spatiotemporal correlations between the electronic and lattice degrees of freedom as a function of doping and temperature. Finally, we close in  Sec.~\ref{Sec:Conclusion} with some additional discussion of our results and our conclusions. 

\begin{figure}[t] 
\psfig{figure=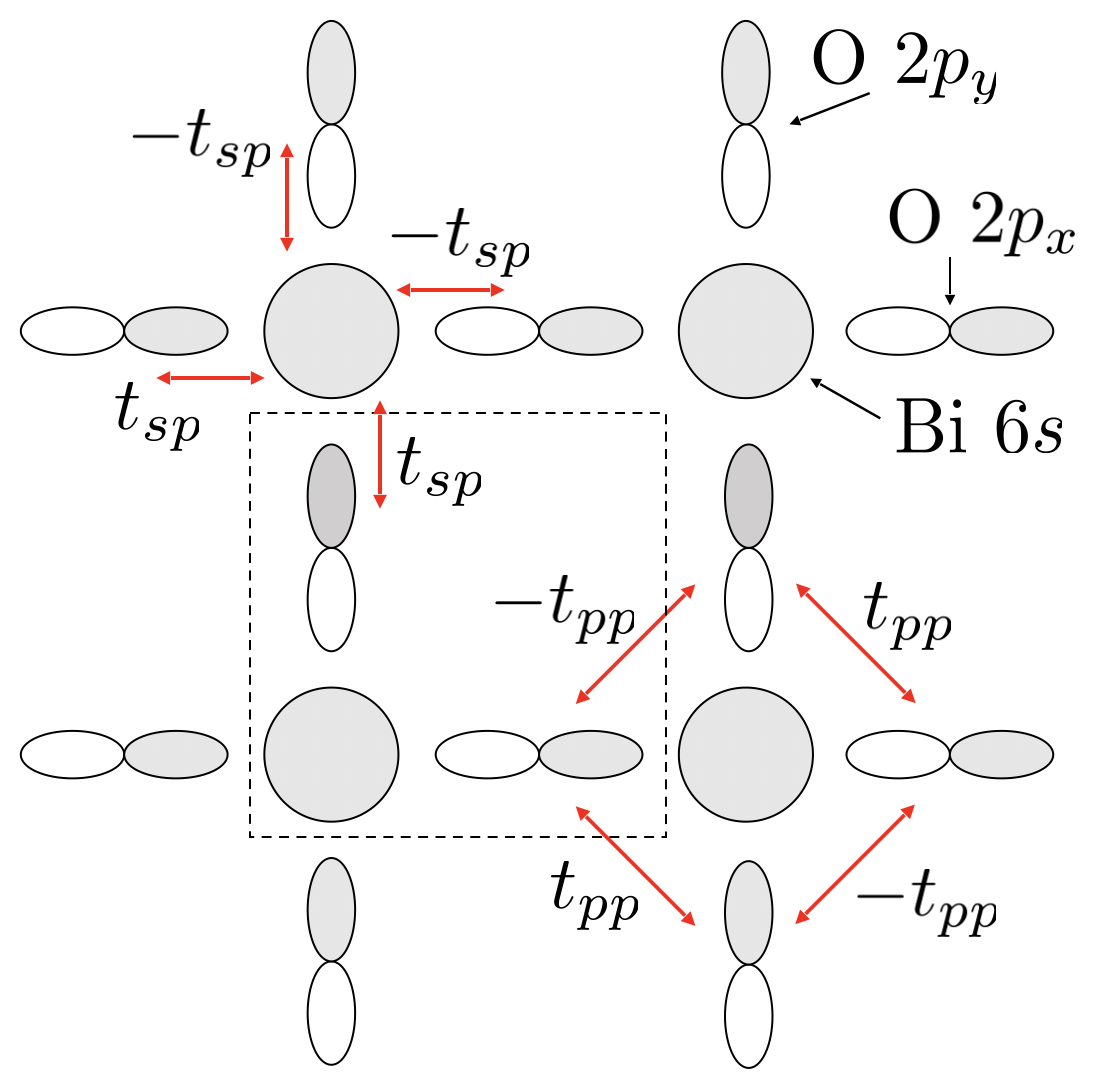, width=.4\textwidth,angle=0,clip}
\caption{(Color online) A sketch of the two-dimensional BiO$_2$ model. The unit cell (indicated by the dashed box) includes the Bi 6s orbital and the O 2$p_x$ and O 2$p_y$ orbitals oriented along the Bi-Bi bonds. The phase convention for the nearest-neighbor Bi-O and O-O hopping integrals is also indicated.}
\label{fig1}
\end{figure}
\section{Model and Methods}
\subsection{The Model}\label{Sec:Model}
We consider a three-orbital model consisting of a Bi $6s$ orbital and two O $2p_\sigma$ orbitals on a two-dimensional (2D) Lieb lattice, as sketched in Fig.~\ref{fig1}. The equilibrium positions of the atoms are given by ${\bf r}+{\boldsymbol{\tau}}_\alpha$, where $\mathbf{r} = r_x {\bf a}_x + r_y {\bf a}_y$ is a lattice vector, 
$r_{x},~r_y\in \mathbb{Z}$, ${\bf a}_x = (a, 0)$ and ${\bf a}_y = (0, a)$ are primitive lattice vectors, $a$ is the equilibrium Bi-Bi bond distance, and ${\boldsymbol{\tau}}_s = (0,0)$, ${\boldsymbol{\tau}}_x = (\tfrac{a}{2},0)$, and ${\boldsymbol{\tau}}_y = (0,\tfrac{a}{2})$, are basis vectors for the $6s$, $2p_x$, and $2p_y$ orbitals, respectively. Following  Ref.~\cite{Steve2020}, the heavier Bi atoms are held stationary while the lighter O atoms are allowed to displace by an amount $X_{{\bf r}\delta}$ ($\delta = x,y$) along the Bi-Bi bond directions and the motion of the oxygen is coupled to the carriers through bond-length dependent Bi-O hopping integrals. We also neglect the kinetic energy of the oxygen vibrations and treat the oxygen displacements as classical variables. 

We emphasis that our treatment is different from conventional frozen phonon approaches in that it captures thermal and spatial fluctuations of lattice displacements. By allowing for both kinds of fluctuations, our model can describe physics beyond Hartree-Fock mean-field theory. For example, this approach correctly captures the non-monotonic dependence of the N{\'e}el temperature for the three-dimensional half-filled single-band Hubbard model~\cite{Mukherjee2014}.  

The model's Hamiltonian is $H=H_\mathrm{el}+H_\mathrm{lat}$, where 
\begin{align} \label{Hel}
H_\mathrm{el} &= \sum_{\substack{\langle \mathbf{r}\delta \rangle \\ \sigma}} \left[ t(X^{\phantom\dagger}_{\mathbf{r}\delta}) s^\dagger_{\mathbf{r}\sigma}p^{\phantom\dagger}_{\mathbf{r} \delta \sigma} +h.c. \right] + t_{pp} \sum_{\substack{\langle \mathbf{r}\delta\delta^\prime \rangle \\ \sigma}} P^{\phantom\dagger}_{\delta\delta'} p^\dagger_{\mathbf{r}\delta \sigma} p^{\phantom\dagger}_{\mathbf{r} \delta' \sigma}\nonumber\\
&+ \sum_{\mathbf{r}\sigma} \left[(\epsilon_s-\mu) \hat{n}^s_{\mathbf{r}\sigma} + (\epsilon_p-\mu) (\hat{n}^{p_x}_{\mathbf{r}\sigma} + \hat{n}^{p_y}_{\mathbf{r}\sigma})\right] 
\end{align}  
describes the electron motion through the lattice for a given set of 
oxygen displacements, and 
\begin{align} \label{Hlat}
H_\mathrm{lat} &=  \sum_{\mathbf{r}} \frac{1}{2} K [X^2_{\mathbf{r}x}+X^2_{\mathbf{r}y}]
    + \sum_{\mathbf{r}} \frac{1}{4} \alpha [X^4_{\mathbf{r}x}+X^4_{\mathbf{r}y}]
\end{align}
describes the potential energy of the ions and contains both harmonic $K$ and anharmonic $\alpha$ contributions. (The anharmonic terms are needed in this case to ensure that reasonable magnitudes for the displacements are obtained for realistic parameter choices.) The operators $s^\dagger_{\mathbf{r}\sigma}$ ($s^{\phantom\dagger}_{\mathbf{r}\sigma}$) and $p^\dagger_{\mathbf{r}\delta\sigma}$ ($p^{\phantom\dagger}_{\mathbf{r}\delta\sigma}$) create (annihilate)  spin-$\sigma$ holes on the Bi $6s$ and O $2p_\delta$ orbitals, respectively; $X_{{\bf r}\alpha}$ is the displacement of O atom $\alpha$ measured relative to its equilibrium position; $\langle ... \rangle$ denotes a sum over the nearest-neighbor orbitals so that $\delta,\delta^\prime=\pm x, \pm y$ index the four O atoms surrounding each Bi; $\epsilon_s$ and $\epsilon_p$ are the Bi and  O  site energies, respectively; $\mu$ is the chemical potential; and $t_{sp}(X_{{\bf r}\delta})$ and $t_{pp}$ are the Bi-O and O-O nearest neighbor hopping integrals, respectively. Here, we neglect the distance dependence of the latter (the magnitude of $t_{pp}$ is small), while the former are determined according to Harrison's rule~\cite{Harrison,Lau2013}
\begin{align} \label{H}
t(X_{\mathbf{r}x}) &= -t_{sp} (1+X_{\mathbf{r}x})^{-2} \nonumber \\
t(X_{\mathbf{r}y}) &= -t_{sp} (1+X_{\mathbf{r}y})^{-2} \nonumber \\
t(X_{\mathbf{r}-\hat{x},x}) &= t_{sp} (1-X_{\mathbf{r}-\hat{x},x})^{-2} \nonumber \\
t(X_{\mathbf{r}-\hat{y},y}) &= t_{sp} (1-X_{\mathbf{r}-\hat{y},y})^{-2}, 
\end{align}
where $t_{sp}$ and $t_{pp}$ are the Bi-O and O-O hopping integrals in the absence of O displacements. The phase factors for O-O hopping are $P_{\pm x,\pm y} = P_{\pm y,\pm x} = -P_{\pm x,\mp y} = -P_{\mp y,\pm x} = 1$, as sketched in Fig~\ref{fig1}. We emphasize that our model invokes neither the linear approximation for the electron-lattice interaction~\cite{Adolphs2013,Li2015} nor the harmonic approximation for the lattice potential but treats the lattice displacements in the adiabatic limit.

Throughout, we work on an $N = N_x \times N_y$ lattice ($3N$ orbitals in total) and in the context of BaBiO$_3$ with $a = 4.34$~$\mathrm{\AA}$ as the unit length. We also fix the phonon energy to $\hbar\Omega = \hbar \sqrt{K/Ma^2} = 30$ meV so that $K \sim 65$ eV for a realistic oxygen mass $M$. The energy of the bond-stretching phonons in bulk BaBiO$_3$ are approximately two times larger; however, we have found that larger values of $\Omega$ suppress bipolaron formation in our semiclassical model and underestimates the average lattice displacements deep in the bond disproportionate state in comparison to experiments. (This discrepancy may be related to neglecting the kinetic energy of the atomic displacements.) To ensure that the average displacements are consistent with the lattice structure of BaBiO$_3$~\cite{Sleight2015}, we set $\alpha$ by imposing  $|X_{\mathbf{r}\delta}|\approx 0.03a$ (or $0.13$ \AA) at half-filling and at low temperatures.  For the remaining parameters, we adopt $t_{sp} = 2.08$, $t_{pp} = 0.056$, $\epsilon_s = 6.42$, and $\epsilon_p = 2.42$ (in units of eV) to facilitate comparisons with Ref.~\cite{Steve2020}. This choice of parameters places the model firmly in the negative charge transfer regime, where the holes preferentially occupy the oxygen sublattice. 

\subsection{Classical Monte Carlo Method}
We studied the model using a combined exact diagonalization (ED) and classical Monte Carlo (MC) method (ED+MC), which is useful for simulating Hamiltonians where classical degree of freedoms (the lattice displacements) couple to quantum ones (the fermions). This approach is very similar to the one used recently to study single- 
and multi-orbital spin-fermion models for the cuprates~\cite{Mukherjee2014, Mostafa2018} and Fe-based~\cite{Bishop2016} superconductors. 

For our classical phonon displacements, the expectation value of an observable $\hat{O}$ in the grand canonical ensemble is given by
\begin{align}\nonumber
\langle \hat{O}\rangle&= \frac{1}{\mathcal{Z}}\mathrm{Tr}\left[e^{-\beta H}\hat{O}\right]\\
&=\frac{1}{\mathcal{Z}}\int dX \sum_{m} e^{-\beta E_m(\{X_{\mathbf{r}\delta}\})}\ket{m}\hat{O}\bra{m}, \label{Eq:expectation_value}
\end{align}
where $\mathcal{Z}$ is the partition function, $\beta \equiv 1/T$ $(k_B = 1)$ is the inverse temperature, $\int dX$ is shorthand for a $2N$-dimensional integral over the atomic displacements, and $\bra{m}$ are the eigenstates of $H$, computed for a given configuration of displacements $\{X_{\mathbf{r}\delta}\}$. 

In the ED+MC method, the multidimensional integral in Eq.~(\ref{Eq:expectation_value}) is evaluated using the conventional Metropolis MC algorithm~\cite{Hastings}. Specifically, in each MC step, updates of the type $X_{\mathbf{r}\delta}\rightarrow X^\prime_{\mathbf{r}\delta} = X_{\mathbf{r}\delta}+\Delta X$ are proposed at each site, where $\Delta X$ is drawn from a uniform probability distribution. These  updates are accepted or rejected with a probability $P_{X\rightarrow X^\prime} = \exp(-\beta \Delta E)$, where $\Delta E$ is the change in the total energy of the system. 
Each MC sweep consists of repeating this process at all lattice sites either serially or randomly. Note that diagonalizing $H$ has a computational cost of $O(N^3)$ so that the computational complexity to perform one full MC sweep scales as $O(N^4)$.

To compute each observable, the desired physical quantities are measured periodically after the sampling procedure has obtained thermodynamic equilibrium. Because of the $O(N^4)$ scaling in ED+MC, we are limited to lattices up to $N=10\times 10$ in size, and most of our results will be for this lattice size. We did, however, also simulate larger lattices by employing the traveling cluster approximation (TCA), which improves the scaling. 

The TCA reduces the computational complexity of standard ED+MC simulations from $O(N^4)$ to $O(N)$~\cite{TCA1984}. TCA scheme relies on an appropriately defined cluster of linear dimension $L_c$ around a particular lattice site where the MC update is attempted. The proposed MC update is accepted or rejected depending on the energy difference obtained by diagonalizing the problem only within the cluster, rather than for the full lattice. Typically, the cluster geometry is chosen to be the same as the full lattice (as done in this work). In a two-dimensional square lattice, for instance, the number of sites in the cluster is $N_c=L^2_c$. Consequently, each update can be accepted or rejected with an computational cost of $O(N^3_c)$ as opposed to $O(N^3)$, where typically $N_c \ll N$. In addition, periodic boundary conditions are imposed on the cluster. Thereby, the cluster acts as an independent ensemble whose equilibrium with the remainder of the full lattice is maintained in a grand canonical framework. Due to the use of periodic boundary conditions, any site within the cluster can be equivalently chosen as the ``update site'', regardless the origin within the cluster. 
TCA scheme has been extensively tested and utilized in various contexts~\cite{TCA1984,TCA2006,TCADagotto,SekharEPL2004}. 
While more advanced implementations such as the parallelized TCA exist~\cite{TCADagotto}, we employ the simplest version of TCA in this paper. 

\subsection{Observables}\label{Sec:Observables}
Our focus is on lattice (bi)polaron correlations and their evolution upon heating and hole doping. The (bi)polarons, where the holes are bound to local breathing distortions of the oxygen sublattice, can be described by the polaron $S_p=\frac{1}{N} \sum_{\mathbf{r}} \langle \hat{p}(\mathbf{r}) \rangle$ and bipolaron $B_p=\frac{1}{N} \sum_{\mathbf{r}} \langle \hat{g}(\mathbf{r}) \rangle$ number operators, where
\begin{equation}\label{Eq:p}
\hat{p}(\mathbf{r})= X_{\mathbf{r} L_s} ( \hat{n}_{\mathbf{r}\uparrow} + \hat{n}_{\mathbf{r}\downarrow} - 2 \hat{n}_{\mathbf{r}\uparrow} \hat{n}_{\mathbf{r}\downarrow})  
\end{equation}
and
\begin{equation}\label{Eq:bp}
\hat{g}(\mathbf{r})= X_{\mathbf{r} L_s} \hat{n}_{\mathbf{r}\uparrow} \hat{n}_{\mathbf{r}\downarrow}.  
\end{equation}
Here,  
$\hat{n}^\pdag_{\mathbf{r}\sigma} = \hat{n}^s_{\mathbf{r}\sigma} + \hat{n}^{L_s}_{\mathbf{r}\sigma}$ and 
$\hat{n}^{L_m}_{\mathbf{r}\sigma}= \sum_{\sigma} L^{\dagger}_{m\mathbf{r}\sigma} L^{\phantom\dagger}_{m\mathbf{r}\sigma}$ are local number operators. The operators $L^{\pdag}_{m\mathbf{r}\sigma}$ ($L^{\dagger}_{m\mathbf{r}\sigma}$) 
with $m=s,d,x,y$ define molecular orbitals from linear combinations of ligand oxygen orbitals surrounding each Bi site~\cite{Steve2020}. They are given by 
\begin{equation}\label{Eq:Ls}
\begin{aligned}
L_{s\mathbf{r}\sigma} &= \frac{1}{2} (p_{\mathbf{r},-x\sigma} + p_{\mathbf{r},-y\sigma}- p_{\mathbf{r}x\sigma} - p_{\mathbf{r}y\sigma}) \nonumber \\ 
L_{d\mathbf{r}\sigma} &= \frac{1}{2} (p_{\mathbf{r},-x\sigma} - p_{\mathbf{r},-y\sigma}- p_{\mathbf{r}x\sigma} + p_{\mathbf{r}y\sigma}) \nonumber \\ 
L_{x\mathbf{r}\sigma} &= \frac{1}{\sqrt{2}} (p_{\mathbf{r},x\sigma} +  p_{\mathbf{r},-x\sigma}) \nonumber \\ 
L_{y\mathbf{r}\sigma} &= \frac{1}{\sqrt{2}} (p_{\mathbf{r},y\sigma} +  p_{\mathbf{r},-y\sigma}).
\end{aligned}
\end{equation}
One can also perform a similar transformation for the oxygen displacements
\begin{align} \label{Xs}
X_{\mathbf{r} L_s} &= \frac{1}{2} (X_{\mathbf{r},-x} + X_{\mathbf{r},-y} - X_{\mathbf{r}x} - X_{\mathbf{r}y} ) \nonumber \\  
X_{\mathbf{r} L_d} &= \frac{1}{2} (X_{\mathbf{r},-x} - X_{\mathbf{r},-y} - X_{\mathbf{r}x} + X_{\mathbf{r}y} ) \nonumber \\  
X_{\mathbf{r} L_x} &= \frac{1}{\sqrt{2}} (X_{\mathbf{r},x} + X_{\mathbf{r},-x}) \nonumber \\  
X_{\mathbf{r} L_y} &= \frac{1}{\sqrt{2}} (X_{\mathbf{r},y} + X_{\mathbf{r},-y}). 
\end{align}

Note that the $L_s$ and $L_d$ operators correspond to the $A_{1g}$ and $E_g$ orbitals in Ref.~\cite{Foyevtsova2015}. Accordingly, in the molecular orbital basis, the optical phonon operator $X_{\mathbf{r},L_s}$ defines the bond disproportionated mode of $A_{1g}$ symmetry, which couples strongly to the carriers in the bismuthates according to DFT calculations~\cite{Arash2018a,Arash2018b}. 
The operator $\hat{p}(\mathbf{r})$ thus measures the combined presence of a single hole on the Bi or $A_{1g}$ molecular orbital located at $\mathbf{r}$ and a local compression of the four surrounding oxygen atoms. The operator $\hat{g}({\bf r})$ measures a similar correlation but involving two holes on 
the BiO$_4$ complex instead of one. 
Also note that we have removed the double occupancy term in the definition of $\hat{p}({\bf r})$, which is already accounted for in the definition of the bipolaron number operator $\hat{g}(\mathbf{r})$. 

When the system has significant bond disproportionation correlations, a two sublattice structure will appear as the oxygen octahedra collapse and expand about alternating Bi sites. This phenomenon will be reflected in a given local quantity $\hat{O}_{\bf r}$ if sublattice averages are performed. In our case, bond disproportionation leads to a bipartite lattice structure, and so we define $A$/$B$ sublattice averages using 
\begin{equation}\label{sublattice_avg}
\langle \hat{O}^{A(B)_{\bf r}}\rangle 
= \frac{2}{N}\sum_{\bf r}   \langle \frac{1}{2}\left[1\pm (-1)^{r_x+r_y}\right]
\hat{O}_{\bf r}\rangle, 
\end{equation}
where the $+$ ($-$) sign corresponds to lattice sites on the $A$ ($B$) sublattice. 

Finally, the total hole concentration $\rho = \tfrac{1}{N}\sum_{{\bf r},\alpha} \langle n_{{\bf r},\alpha} \rangle$ is determined from a sum over all orbitals in the cluster. 

\section{Results}\label{Sec:Results}
\subsection{Insulating behavior at half-filling}\label{Sec:Insulating}
\begin{figure}
    \centering
    \psfig{figure=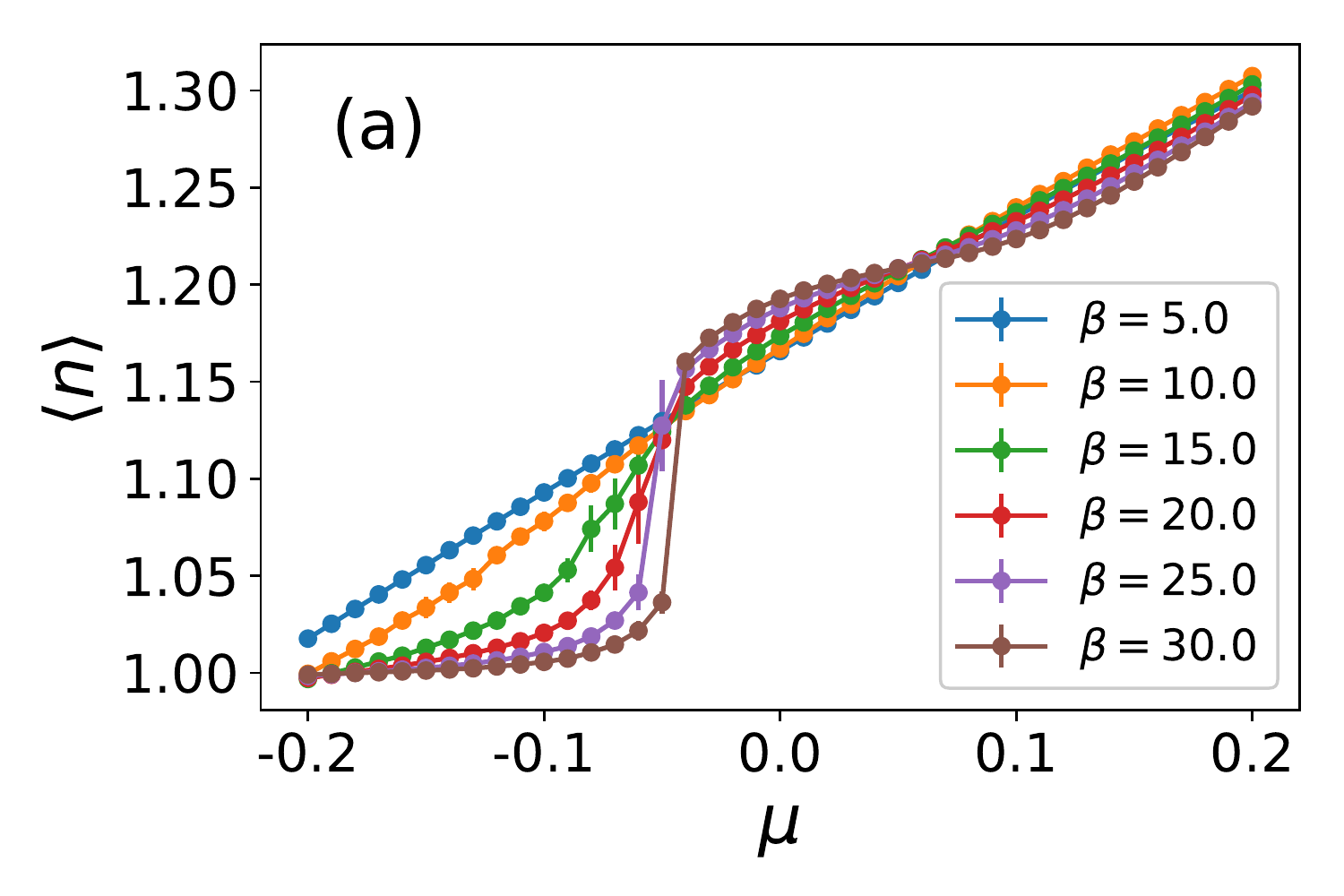,width=.45\textwidth,angle=0,clip=true,trim = 0.0cm 1.4cm 0.0cm 0.0cm}
\psfig{figure=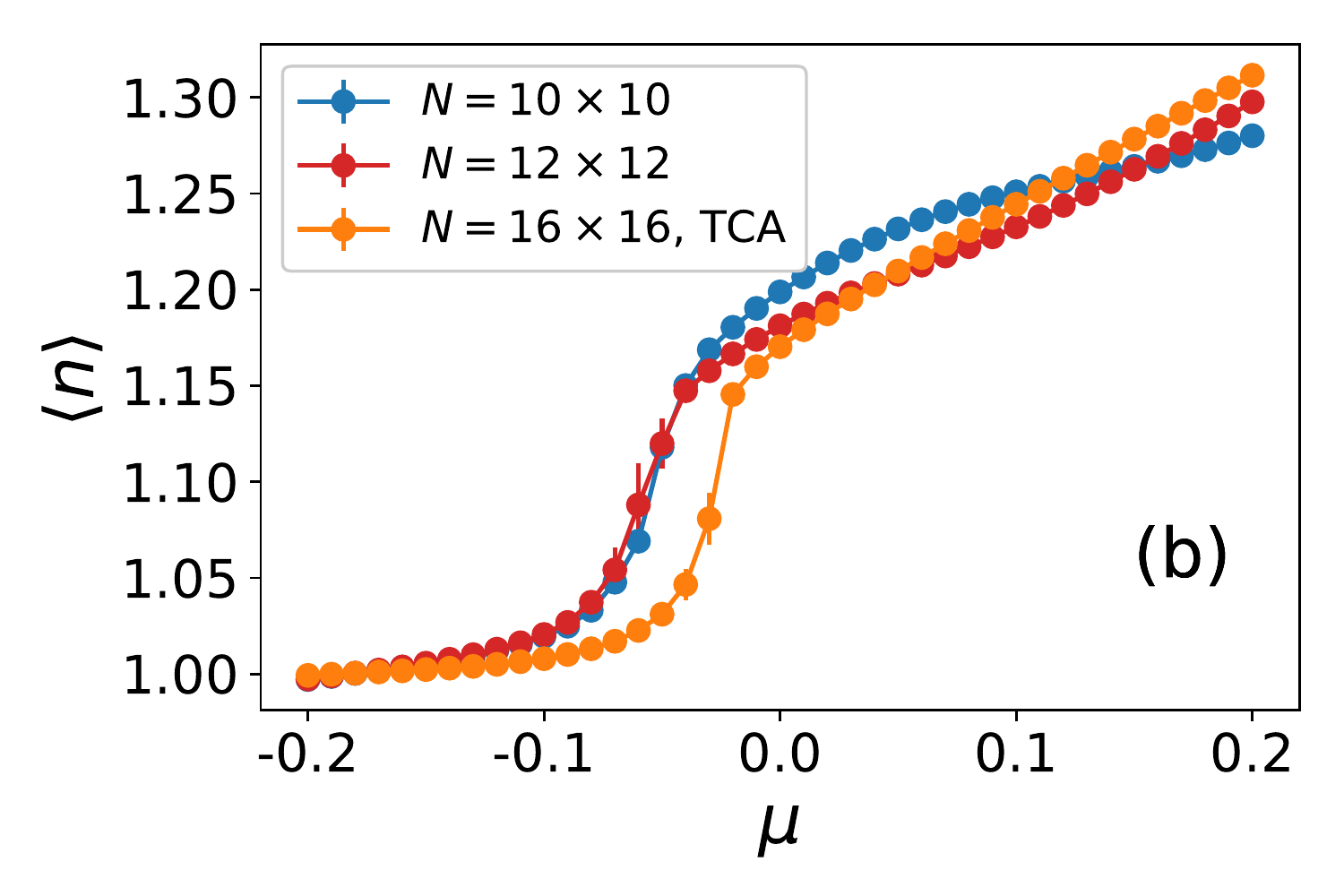,width=.45\textwidth,angle=0,clip} 
    \caption{(a) The average fillings of a $N=12\times 12$ cluster as a function of the chemical potential $\mu$ for various temperatures; (b) Variation of the average filling vs $\mu$ for different lattice sizes at $\beta=20$ eV$^{-1}$.} 
    \label{Fig:N_vs_mu}
\end{figure}
Our first task is to establish the existence of a metal-to-insulator transition at half-filling, which will form the basis for our discussion in the remainder of the paper. 

Figure~\ref{Fig:N_vs_mu}(a) plots the average hole concentration $\langle n \rangle$ as a function the chemical potential $\mu$ and temperature. At high temperature, $\langle n \rangle$ varies monotonically with $\mu$, which is indicative of a metallic system with a non-zero compressibility. As we lower the temperature, however, the filling begins to form a plateau at a function of $\mu$ around $\langle n \rangle= 1$ holes/Bi, consistent with the opening of a gap and the transition to an insulating state. We also see some hints of a second plateau around $\mu=0.075$ and $\langle n \rangle \approx 1.2$ holes/Bi, which is most likely due to finite-size effects. This hypothesis is confirmed in Fig.~\ref{Fig:N_vs_mu}(b), which plots the filling vs chemical potential as a function of cluster size and at a fixed $\beta = 20$ eV$^{-1}$. Here, we observe that the slope of the $\langle n \rangle$ vs $\mu$ curve around $\mu = 0.075$ increases on larger lattices while the plateau at half-filling is robust. These results confirm that the half-filled system is an insulator at low temperatures. 

\subsection{Temperature and doping evolution of the local orbital occupations}\label{Sec:Results_Local}
\begin{figure}[b!] 
\psfig{figure=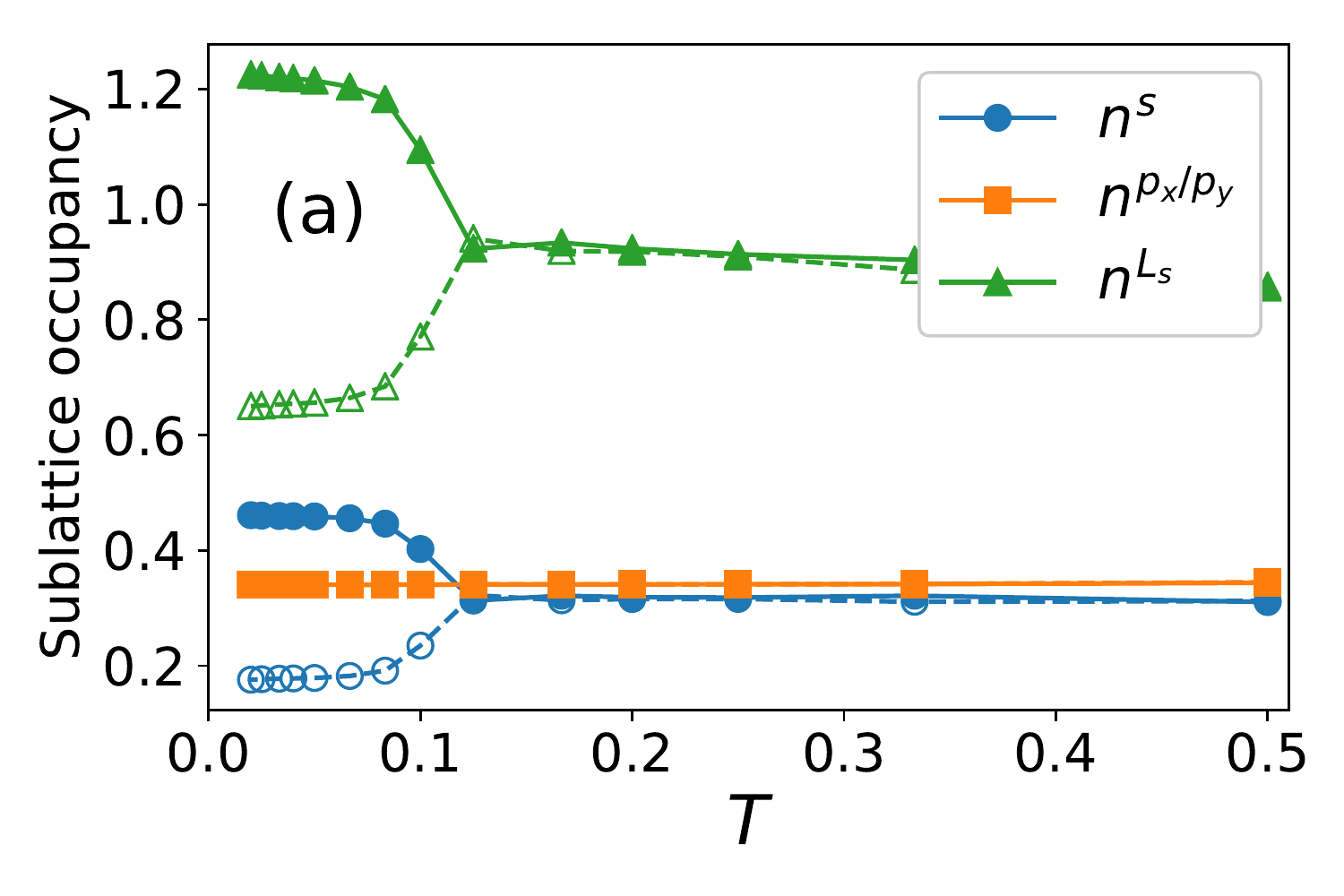,width=.45\textwidth,angle=0,clip=true,trim = 0.0cm 0.6cm 0.0cm 0.0cm}
\psfig{figure=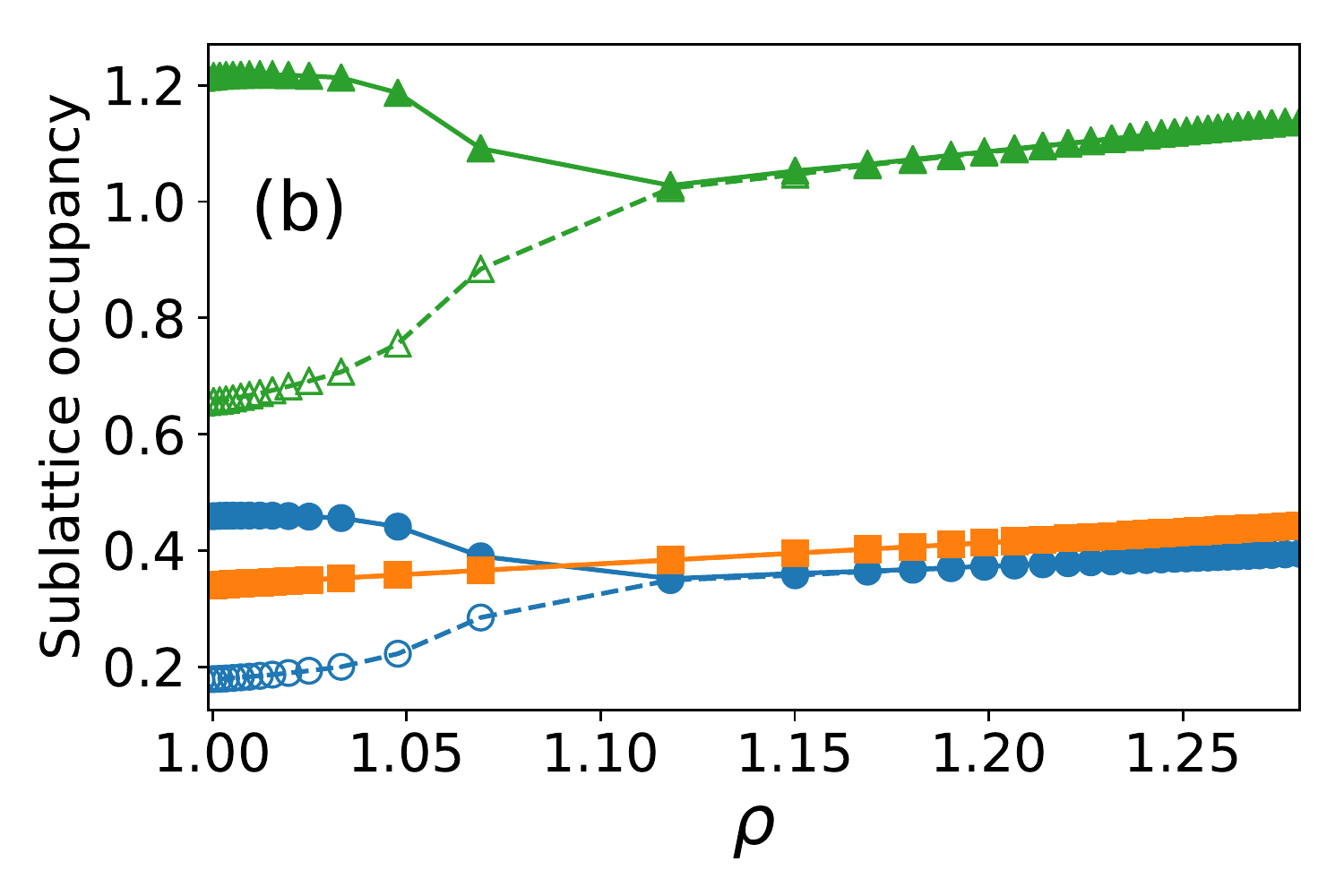,width=.45\textwidth,angle=0,clip} 
\caption{(Color online) Various local quantities obtained by averaging over the two sublattices vs (a) $T$ at half-filling $\rho=1$ and (b) $\rho$ at fixed $\beta=20$. The bifurcation at low temperatures and low doping levels indicates signals the system's  tendency towards bond-disproportionation. Heating and/or doping both induce the transition from the bond-disproportionated state towards a uniform phase that can be best characterized as a (bi)polaron-liquid-like phase (see below). These results were obtained on an $10\times 10$ cluster.}
\label{sublatT}
\end{figure}

Next, we examine several of the system's sublattice orbital occupations as a function of temperature $T$ and filling. Fig.~\ref{sublatT}(a) shows results for the average occupancy of the Bi 6s ($n^s$) and O $2p_\delta$ ($n^{p_\delta}$, $\delta = x,y$)   orbitals at half-filling ($\rho = 1$), as well as the $L_s$ molecular orbital ($n^{L_s}$), as a function of temperature. The doping evolution of the same quantities at fixed inverse temperature $\beta = 20$~eV$^{-1}$ is shown in Fig.~\ref{sublatT}(b). The two sublattice averages are plotted using solid and open symbols, respectively, in both panels.

Figure~\ref{sublatT} reveals that there is a significant tendency towards bond-disproportionation approaching half-filling, which is evident in the bifurcation of $n^{s}$ and $n^{L_s}$ at low temperatures. The bifurcation reflects the fact that the $L_s$ molecular orbital hybridizes with the central $6s$ orbital as the O atoms compress around a Bi atom. The hole occupation on the short-bond BiO$_4$ plaquette should, therefore, increase at the expense of hole occupation on the neighboring plaquettes. Increasing the simulation temperature or hole concentration reduces the size of the bifurcation, before it ultimately disappears for $T > 0.125$ eV or a hole density $\rho > 1.15$. [The precises values depend somewhat on the cluster size, see e.g. Fig. \ref{SpBpN}(b).] This behavior reflects a temperature- and/or doping-driven transition from a bond-disproportionated state to one that is uniform once averaged over the cluster. Later, we will show that the ``uniform" phase is, in fact, inhomogeneous when viewed on smaller lengths scales, consistent with the DQMC results~\cite{Steve2020}. We also note that we do not observe any bifurcation of the O $2p_{x,y}$ occupancy, which is expected since all of the O orbitals are equivalent in both the ``uniform" and bond-disproportionated phases~\cite{Millis2013,Johnston2014}. 

The bifurcation in the orbital occupations indicate that strong deviations occur between the compressed and expanded oxygen plaquettes due to the bond-disproportionation. Moreover, the similarity between the behavior of the system upon either heating or doping implies that the mechanisms underlying the temperature- or doping-driven transitions from the bond-disproportionated state to the non-disproportionated one have a common origin. We will return to this point shortly.

\subsection{The formation of lattice (bi)polarons}\label{Sec:Results_Bipolaron}
\begin{figure}[h!] 
\psfig{figure=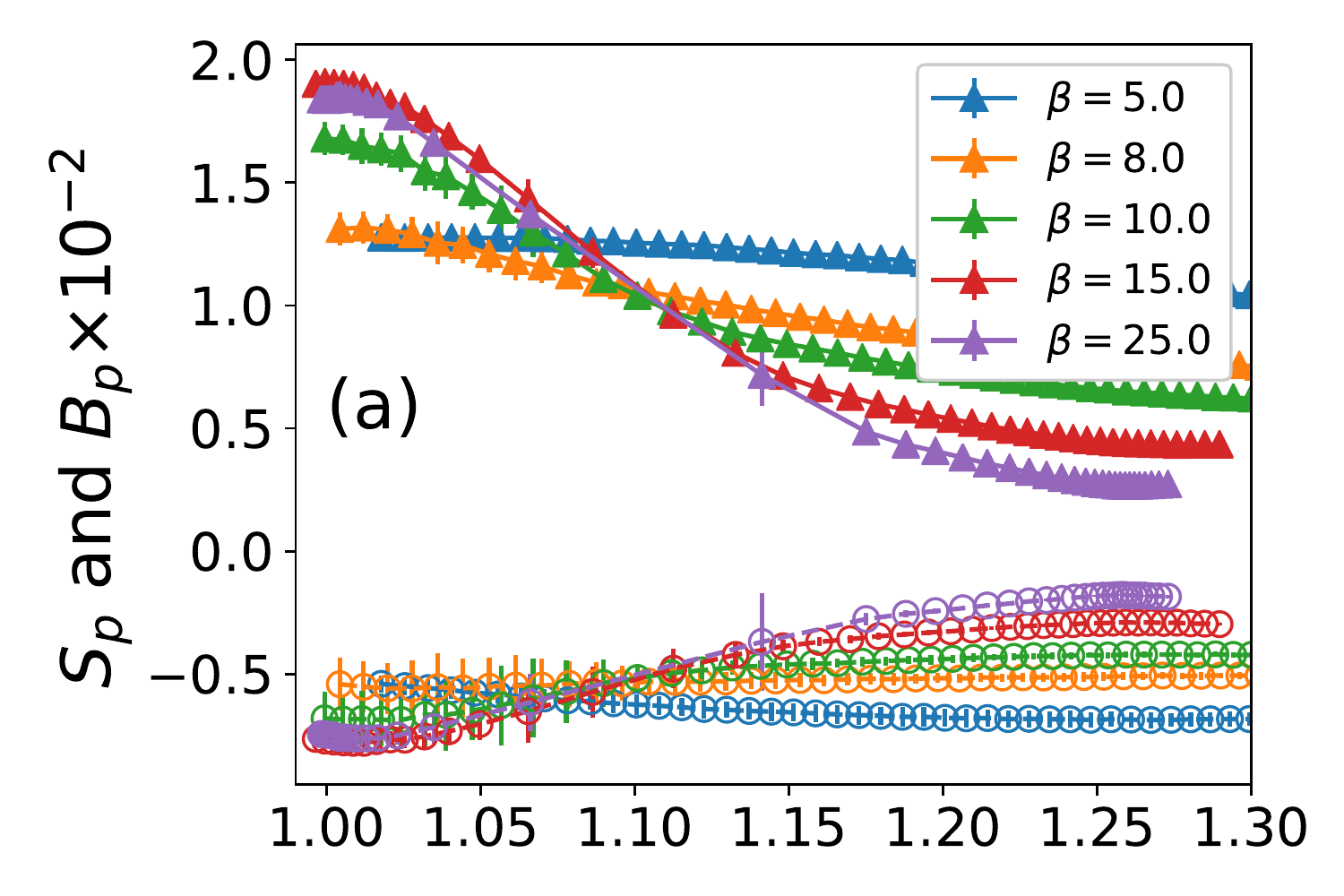,width=.45\textwidth,angle=0,clip=true,trim = 0.0cm 0.5cm 0.0cm 0.0cm} \\
\psfig{figure=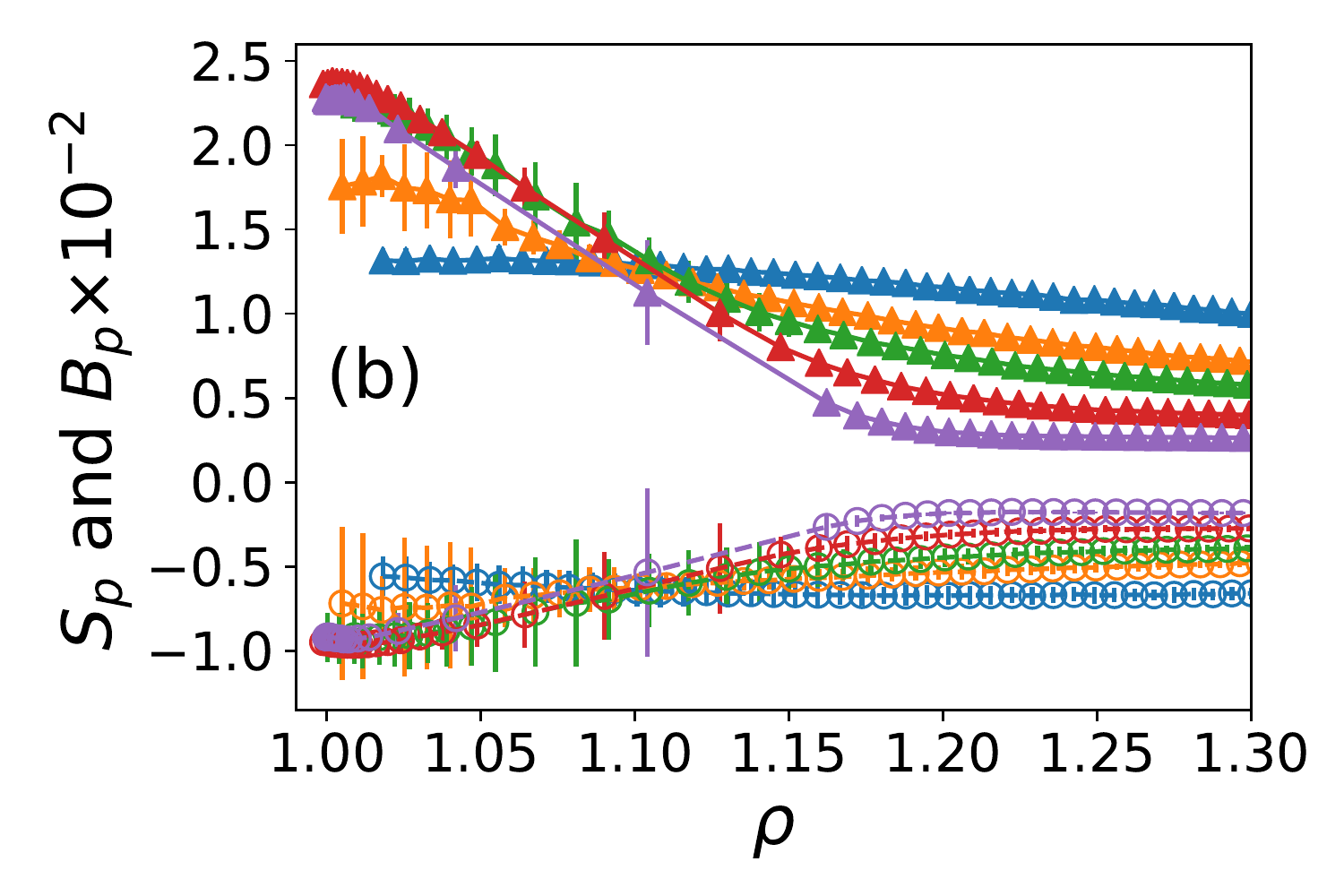,width=.45\textwidth,angle=0,clip} 
\caption{(Color online) Doping evolution of the polaron $S_p$  and  bipolaron $B_p$ numbers versus hole density at various temperatures.  Results are shown for (a) a full calculation of an $N=10 \times 10$ lattice, and (b) for employing the travelling cluster approximation (TCA) with cluster size $N_c=4\times 4$ on  an $N=16 \times 16$  lattice. The phonon energy in both cases was fixed to $\Omega=30$ meV.}
\label{SpBpden}
\end{figure}

\begin{figure}[h!] 
\psfig{figure=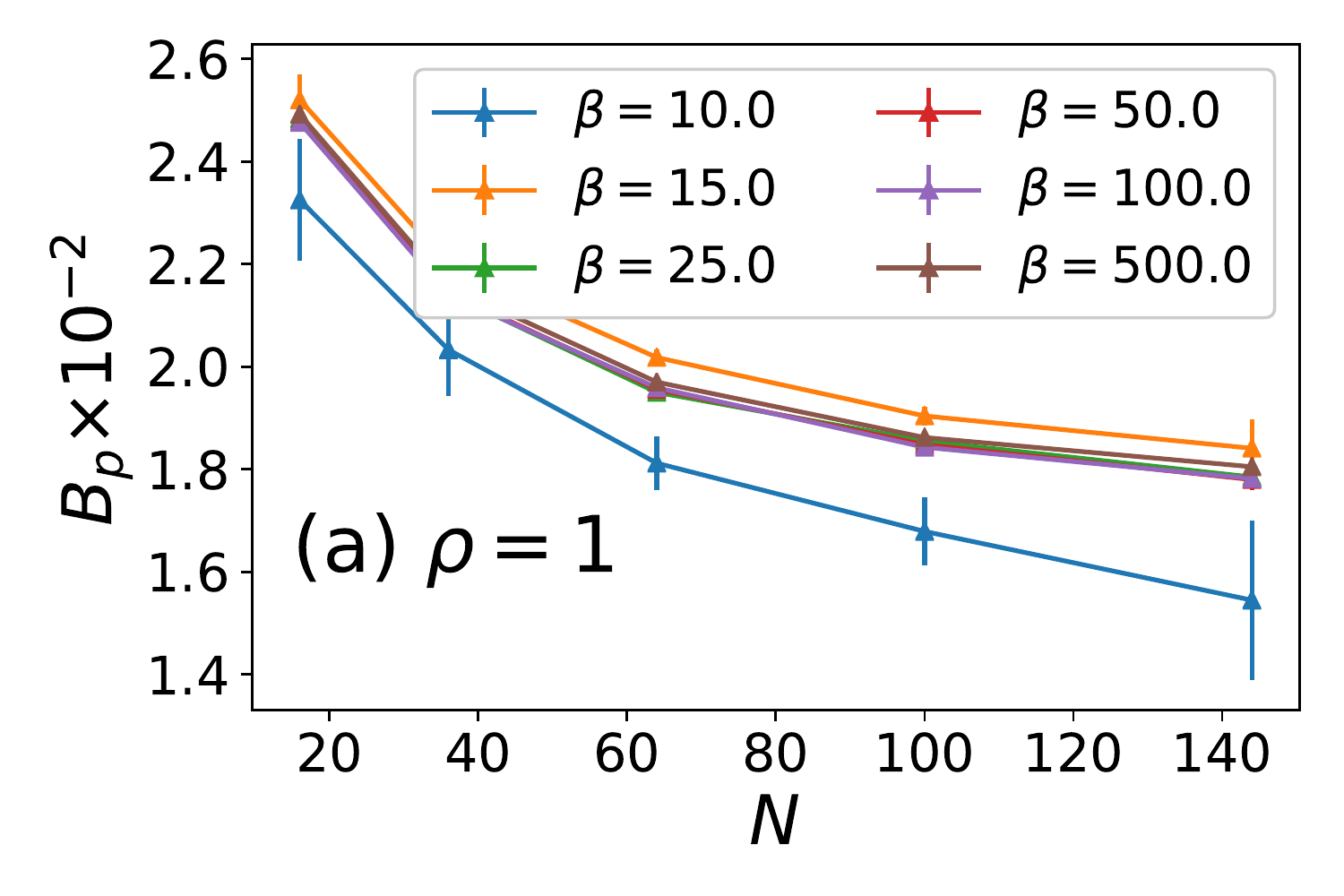,width=.45\textwidth,angle=0,clip=true,trim = 0.0cm 0.5cm 0.0cm 0.0cm} 
\psfig{figure=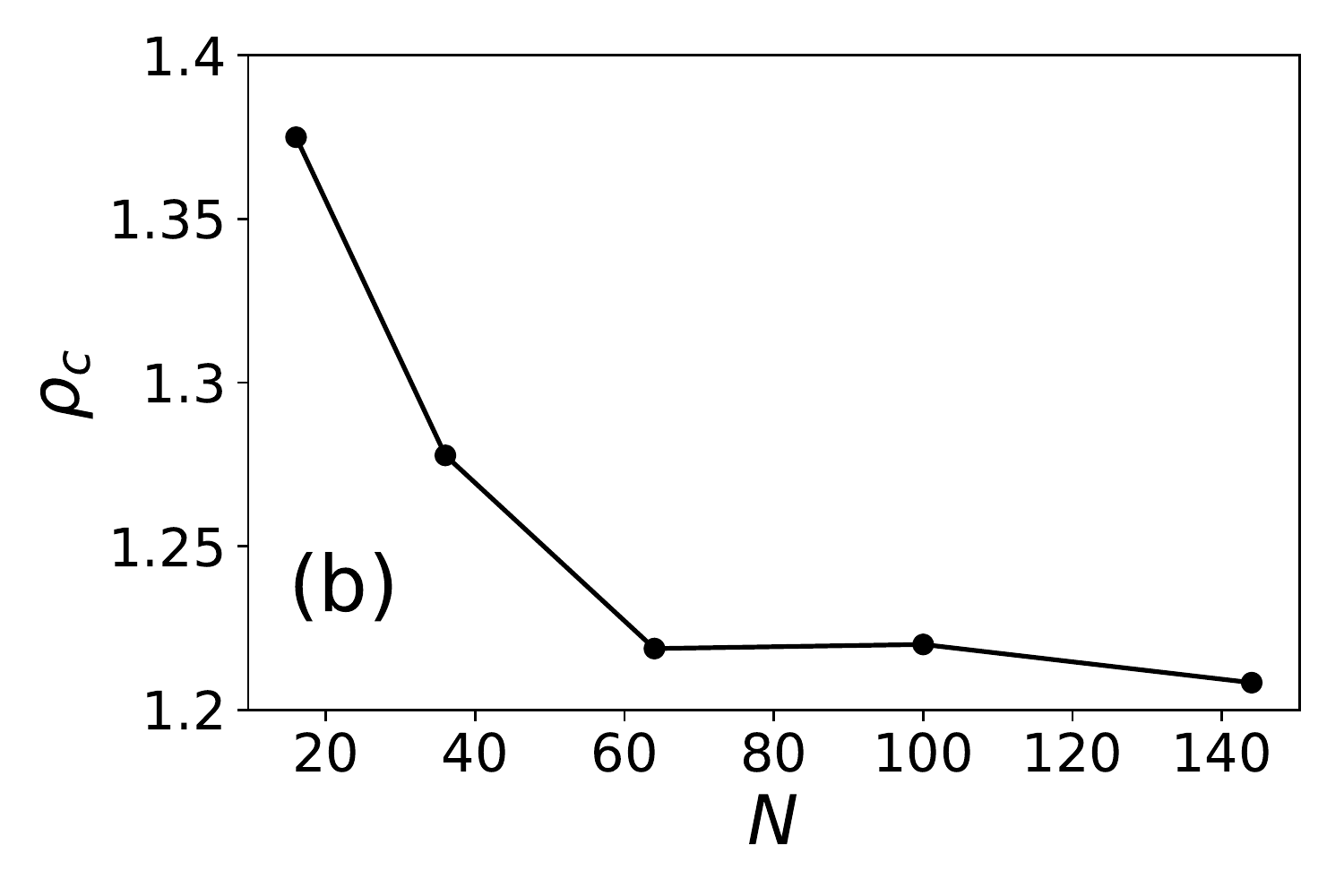,width=.45\textwidth,angle=0,clip} 
\caption{(Color online) A finite-size scaling analysis of the results obtained in Fig.~\ref{SpBpden}. Panel (a) shows the scaling of $B_p$ at half-filling as a function of the total cluster size $N=N_x\times N_y$ at various temperatures. Panel (b) shows the scaling of the critical density $\rho_c$, which is defined as the doping at which $B_p$ vanishes in our full calculation. }
\label{SpBpN}
\end{figure}

Several works~\cite{Siebold1993,FranchiniPRL2009,Naamneh2018,Shamblin2018}, including the previous DQMC study of a comparable model~\cite{Steve2020}, have proposed that the metallic phase of the bismuthates (and other negative CT systems like the Nickelates) can be viewed as a polaron liquid. With this idea in mind, Fig.~\ref{SpBpden} presents the temperature and doping dependence of the average polaron $S_p$ (open symbols) and bipolaron $B_p$ (closed symbols) numbers versus hole density at varying temperatures. Here, we show results for a calculation involving the full diagonalization of a $N=10 \times 10$ cluster [Fig.~\ref{SpBpden}(a)], and for a calculation employing the TCA ($N_c=4\times 4$) on a larger $N=16 \times 16$ cluster [Fig.~\ref{SpBpden}(b)]. The results of both calculations are qualitatively similar; however, we do observe some quantitative differences, indicating that some finite size effects are present. Nevertheless, the qualitative picture is the same. In both cases, we find that there is a significant number of bipolarons in the system at half-filling, and that their number decreases with increasing hole concentrations $\rho>1$. This behavior is more pronounced at low-temperatures, but persists at higher $T$, albeit with a smoother $\rho$ dependence. Moreover, at low $T$, the bipolaron number approaches a small minimum value as the hole concentration increases towards a ``critical" value $\rho_c$, and for $\rho > \rho_c$, $B_p$ remains constant. The doping evolution of the single-polaron operator $S_p$ is qualitatively similar; however, the magnitude of $S_p$ is much smaller than $B_p$, which suggests that the holes tend to collect on the compressed plaquettes in pairs rather than individually. This tendency may have implications for superconductivity in this system.   

The results shown in Fig.~\ref{SpBpden} are qualitatively consistent with those reported in Ref.~\cite{Steve2020} but with some notable differences. The biggest one is that we observe a more rapid decrease in the (bi)polaron numbers with doping at low temperature. This discrepancy may be due to differences in the model parameters, differences in the cluster size, differences arising from the quantum vs classical treatment of the model, or some combination thereof. For example, our model parameters and classical MC treatment places our system squarely in the adiabatic regime $\Omega/E_\mathrm{F} \ll 1$. By contrast, Ref. \cite{Steve2020} worked in the antiadiabatic regime due to technical issues related to the autocorrelation time. This difference means that the polaronic effects are likely more prominent in Ref. \cite{Steve2020} and possibly under predicted here. Comparative studies of the two approaches are needed to clarify this issue. 

\begin{figure*}[t] 
\psfig{figure=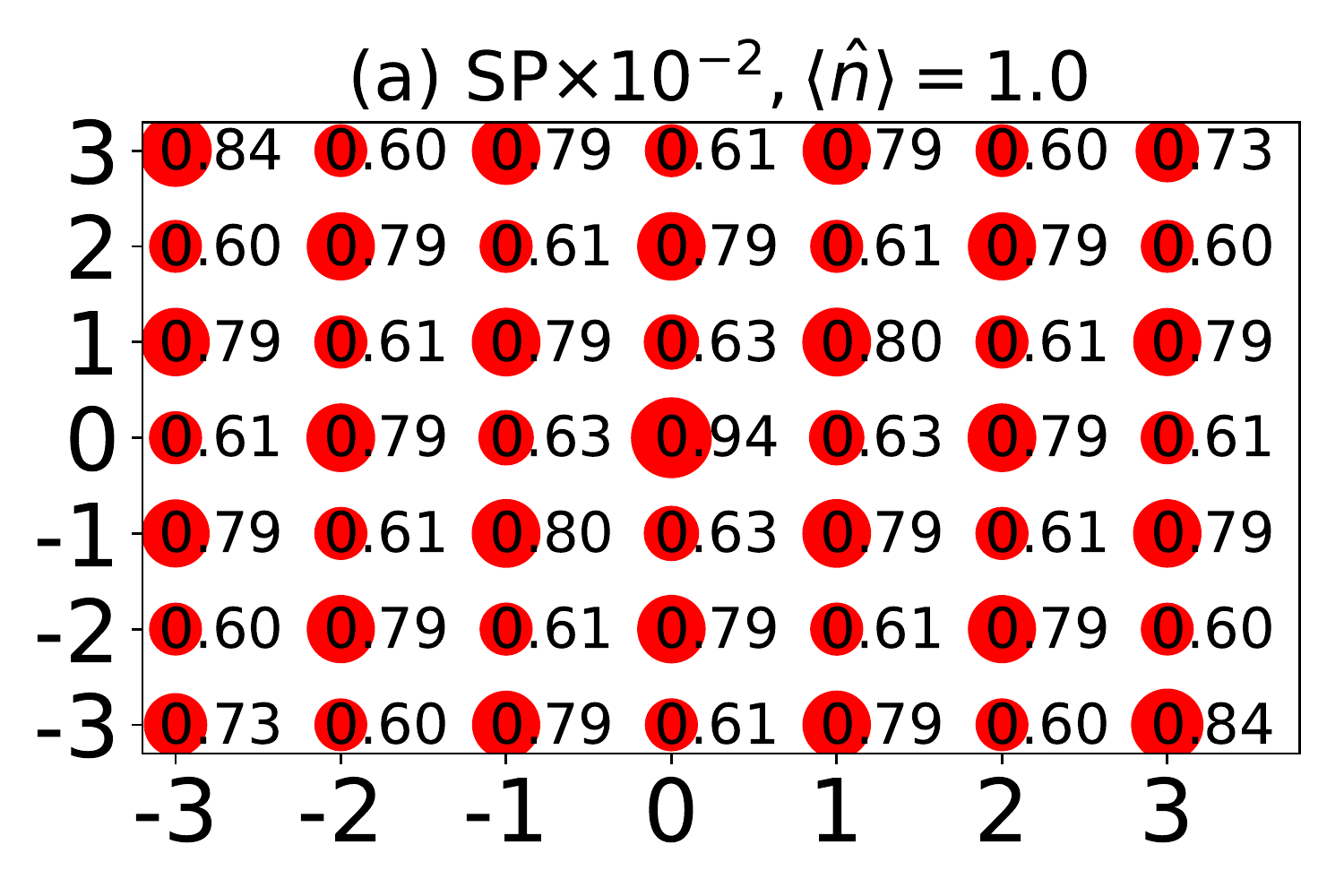,height=6cm,width=.40\textwidth,angle=0,clip=true,trim = 0.0cm 0.0cm 0.0cm 0.0cm} 
\psfig{figure=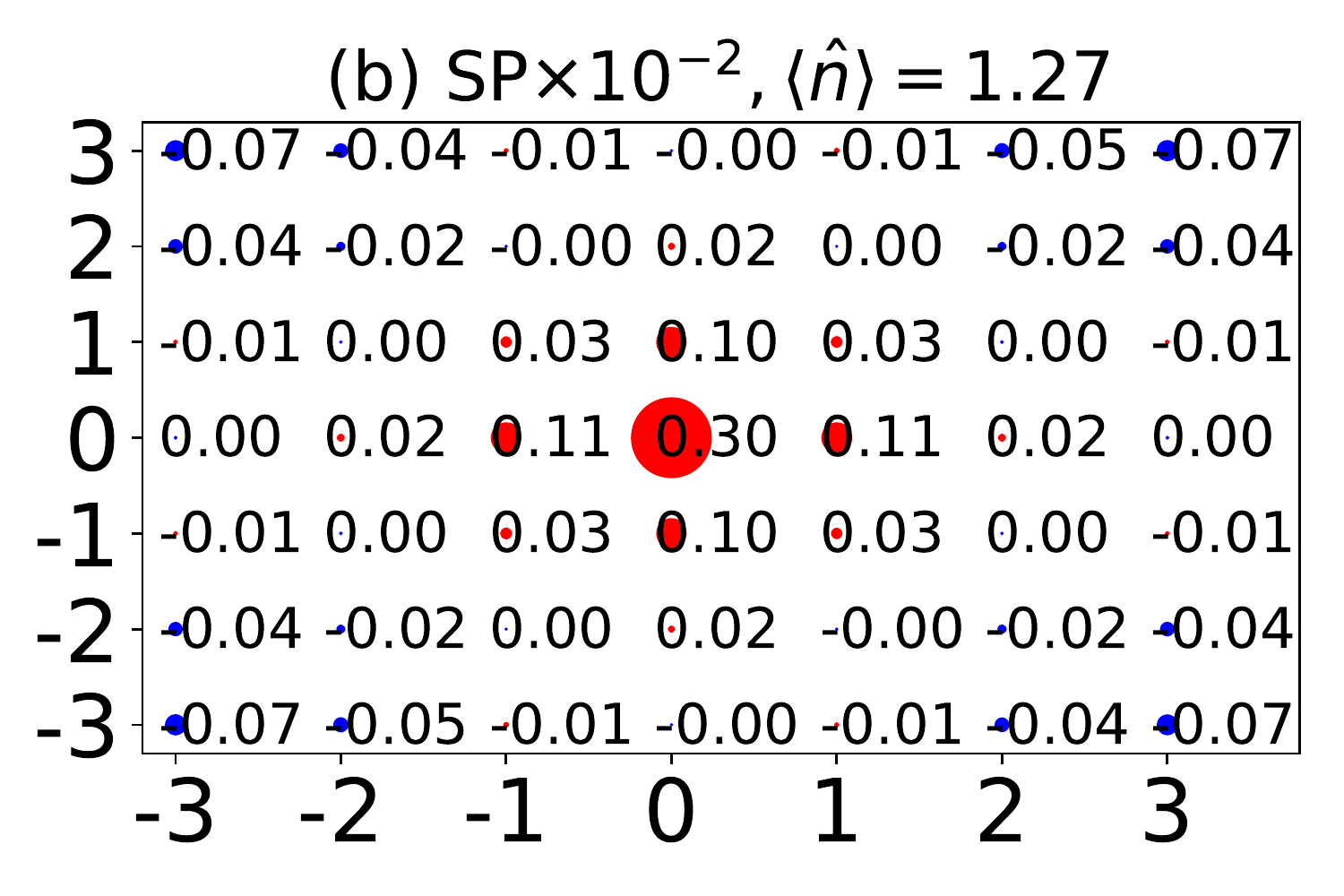,height=6cm,width=.4\textwidth,angle=0,clip} 
\psfig{figure=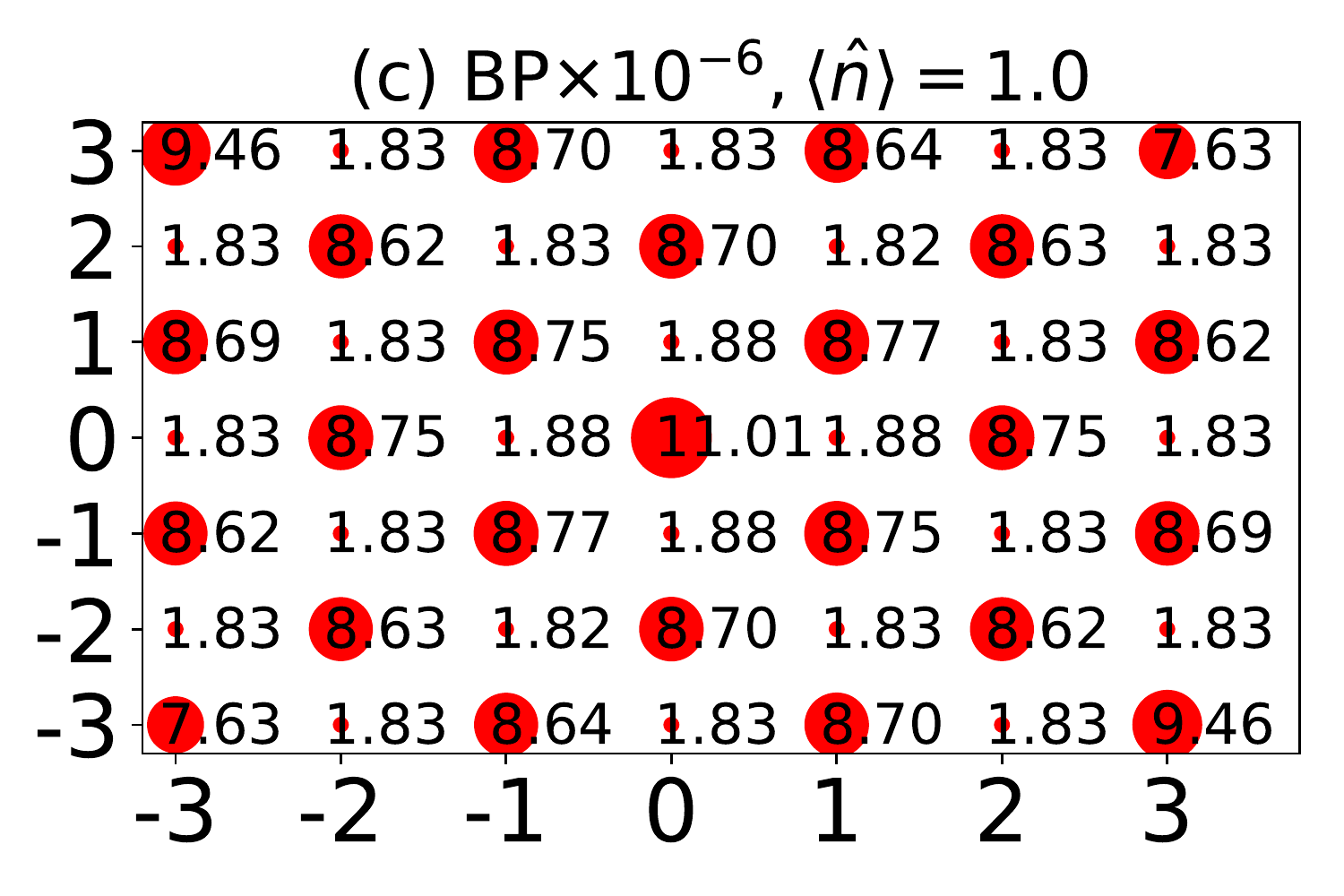,height=6cm,width=.4\textwidth,angle=0,clip=true,trim = 0.0cm 0.0cm 0.0cm 0.0cm} 
\psfig{figure=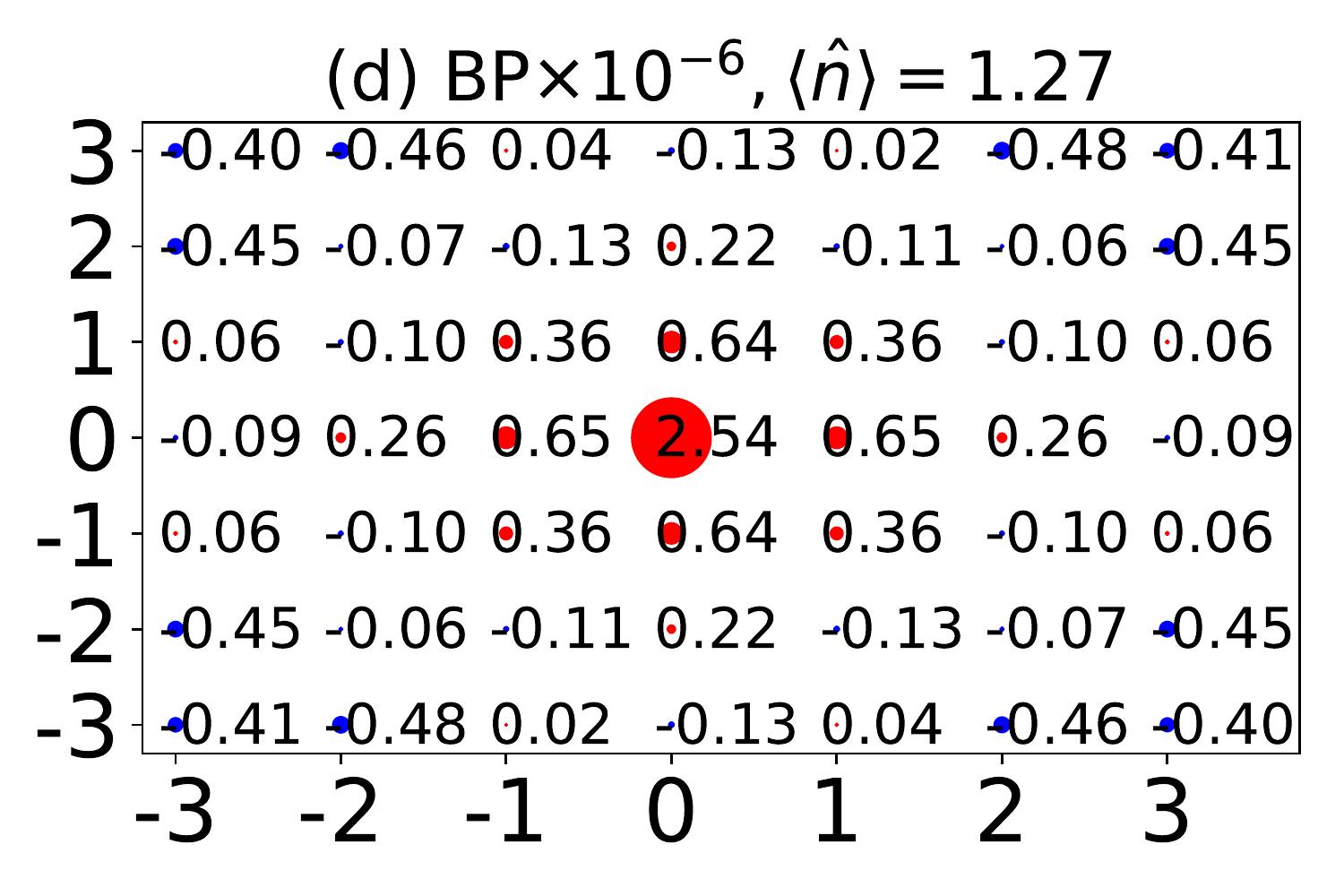,height=6cm,width=.4\textwidth,angle=0,clip} 
\caption{(Color online) The staggered (a-b) polaron $C_\mathrm{SP} (\brp-\br)$ and (c-d) bipolaron $C_\mathrm{BP} (\brp-\br)$ correlation functions as a function of $\brp-\br$. 
Results are plotted here for half-filling ($\rho=1.0$, left column) and for a large hole doping ($\rho=1.27$, right column). All results  were obtained on an $8\times 8$ cluster and at  an inverse temperature of $\beta=25~\mathrm{eV}^{-1}$.  The size of the dots is proportional to the value of the correlation function, while a red (blue) color indicates values larger (smaller) than zero. For reference, the exact numerical values of the correlation functions at each point are also provided. 
}
\label{Fig:Cr}
\end{figure*}

To assess the finite-size effects, Fig.~\ref{SpBpN}(a) plots the evolution of $B_p$ at half-filling as a function of the cluster size $N$, obtained here using full cluster diagonalization. The convergence of $B_p$ for large $N$ is readily apparent, particularly at lower temperatures, demonstrating that the bond disproportionation correlations survives in the thermodynamic limit. 
The critical doping $\rho_c$, where the bipolaron number nearly vanishes, also exhibits some cluster size dependence. For example, Fig.~\ref{SpBpN}(b) shows the evolution of $\rho_c$ with increasing $N$, where we find that saturation at $\rho_c \sim 1.2$ only for $N > 60$. We have observed comparable finite size effects in other quantities of interest, indicating that the quantitative values of the observable here can depend on the cluster size. Nevertheless, Fig.~\ref{SpBpN} assures us that the qualitative physics discussed here is robust against the system size. These caveats should  be kept in mind when interpreting these results. 

We end this subsection with an examination of the real-space structure of the staggered polaron and bipolaron static correlations functions, which are defined as $C_\mathrm{SP} (\brp-\br) = \langle \phi(\brp-\br)
\hat{\rho}(\brp)\hat{\rho}(\br)\rangle$ and $C_\mathrm{BP} (\brp-\br) = \langle \phi(\brp-\br)
\hat{g}(\brp)\hat{g}(\br)\rangle$, respectively, where $\hat{\rho}(\br)$ and $\hat{g}(\br)$ are defined in Sec. \ref{Sec:Observables} and 
$\phi(\br) = (-1)^{(r_x+r_y)}$ reflects an underlying checkerboard-like structure in the correlations. We note here that evaluating $C_\mathrm{SP/BP}$ requires four nested loops over the system's eigenstates, and is, therefore, quite expensive to evaluate. We, therefore, focus on a $N = 8\times 8$ cluster and an inverse temperature of $\beta = 25~\mathrm{eV}^{-1}$. As we will see, this is sufficient to contrast the behavior of the undoped and doped system but finite size effects may be present. 

Figure~\ref{Fig:Cr} plots our results. Panels (a) and (c) show results for the staggered polaron and bipolaron correlation functions at half-filling, respectively. At this filling, both $C_\mathrm{SP}$ and $C_\mathrm{BP}$ develop a real space structure indicative of a bond disproportionate structure and a density modulation on the Bi $6s$ orbitals. In both cases, the staggered correlation functions are positive at all distances. 
Both correlations functions also have a ${\bf Q} = (\pi,\pi)$ modulation in their numerical values, which is more pronounced in the case of the bipolaron correlation function. All of this behavior arises from the  compression/expansion of the ligand O atoms about alternating Bi sites and a consequent weak charge modulation on the Bi orbitals (see Fig. \ref{sublatT}(a) and Ref.~\cite{Steve2020}). Both correlation functions approach large, non-zero values on the longest distances accessible in the cluster, indicating that these correlations are ``long-ranged" from the perspective of the cluster. 

Figures~\ref{Fig:Cr}(b) and \ref{Fig:Cr}(d) show similar results for a high hole concentration $\langle n\rangle = 1.27$. At this doping level, the correlations are significantly reduced in magnitude and rapidly fall off at increasing distances. 
Nevertheless, we observe weak staggered correlations in both $C_\mathrm{SC}$ and $C_\mathrm{BP}$, which suggests that the carriers and their lattice distortions retain some degree of correlation on short length scales. 

\subsection{Fingerprints of a bipolaron liquid at high-temperatures}\label{Sec:Results_Histo}
The results presented so far suggest that the system transitions from an insulating bipolaron lattice, characterized by a bond-disproportionate structure, into a more uniform phase with weak, short-range correlations via heating or doping. But we have only considered the lattice/sublattice averages of various quantities to this point. To investigate the melting process and nature of the uniform phase, we now examine the spatial and temporal correlations in the lattice displacements. To this end, we present detailed histograms of various local physical quantities, collected at each Monte Carlo step of the simulations. These plots provide more detailed snap-shots of the correlations present at each Monte Carlo step and thus supply additional information on the microscopic relationships between several observable quantities and the specific lattice configurations. 

\begin{figure}[t] 
\psfig{figure=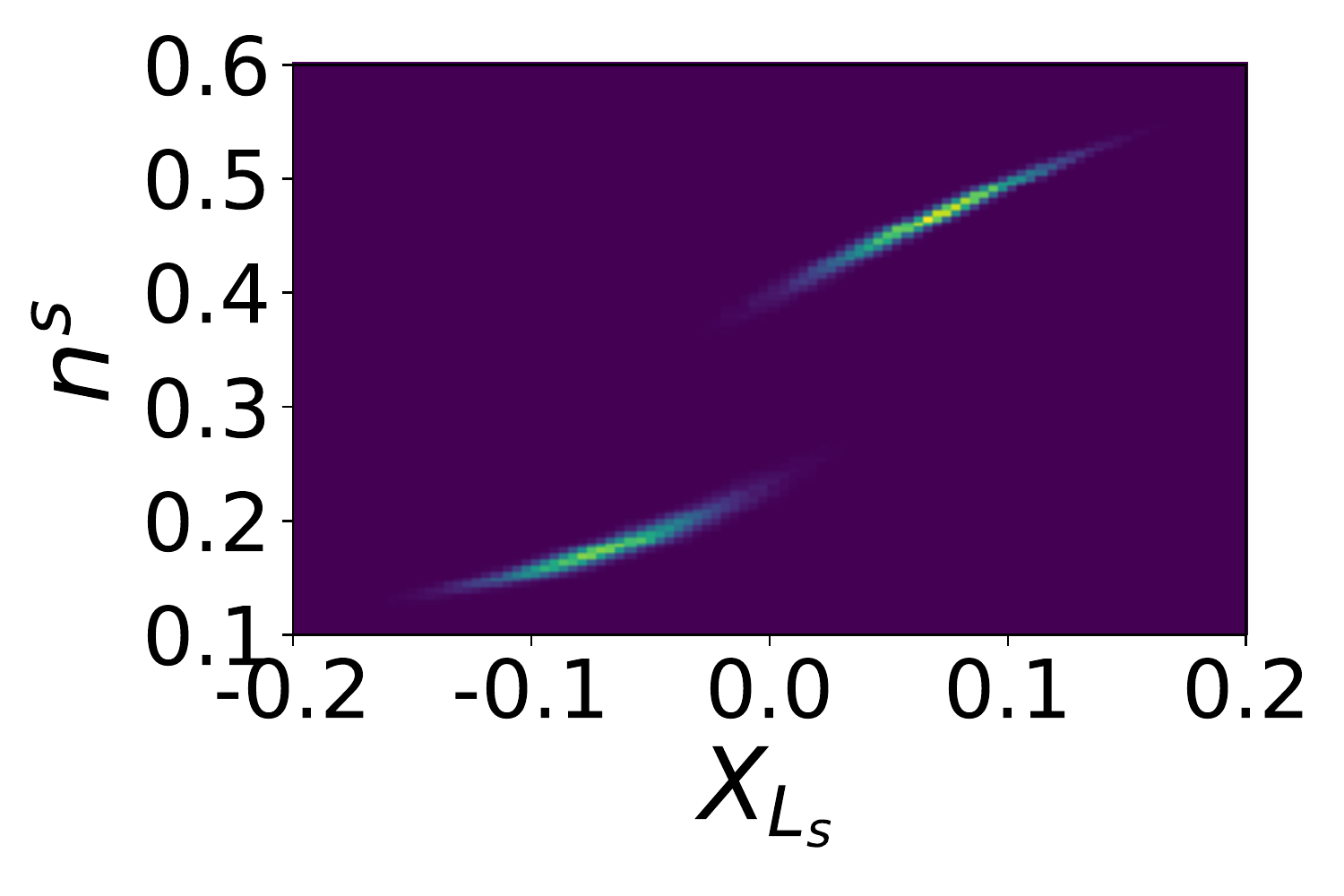,
height=2.2cm,width=2.8cm,angle=0,clip=true,trim = 0.0cm 0.5cm 0.4cm 0.0cm} 
\psfig{figure=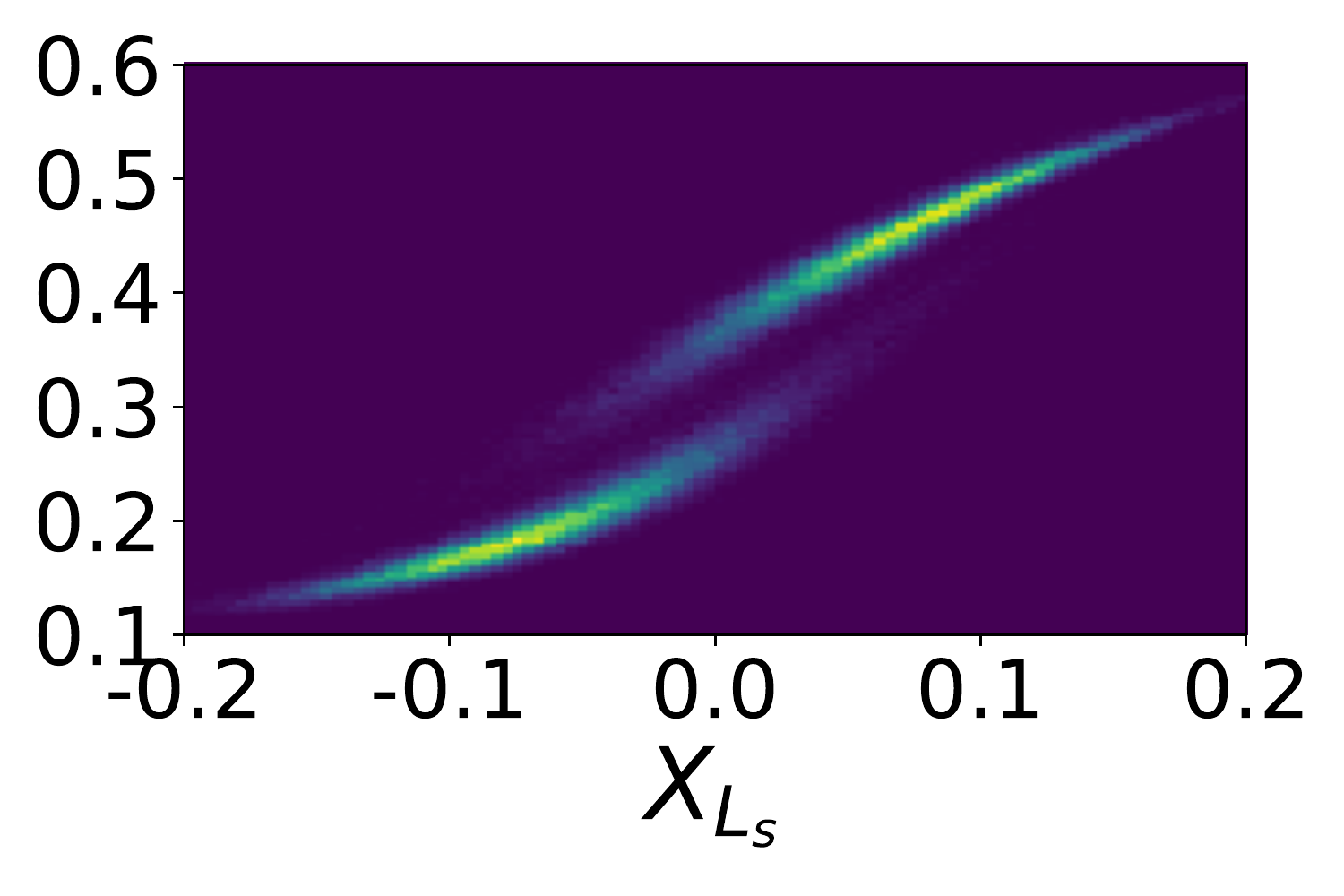,
height=2.2cm,width=2.8cm,angle=0,clip=true,trim = 0.0cm 0.5cm 0.4cm 0.0cm} 
\psfig{figure=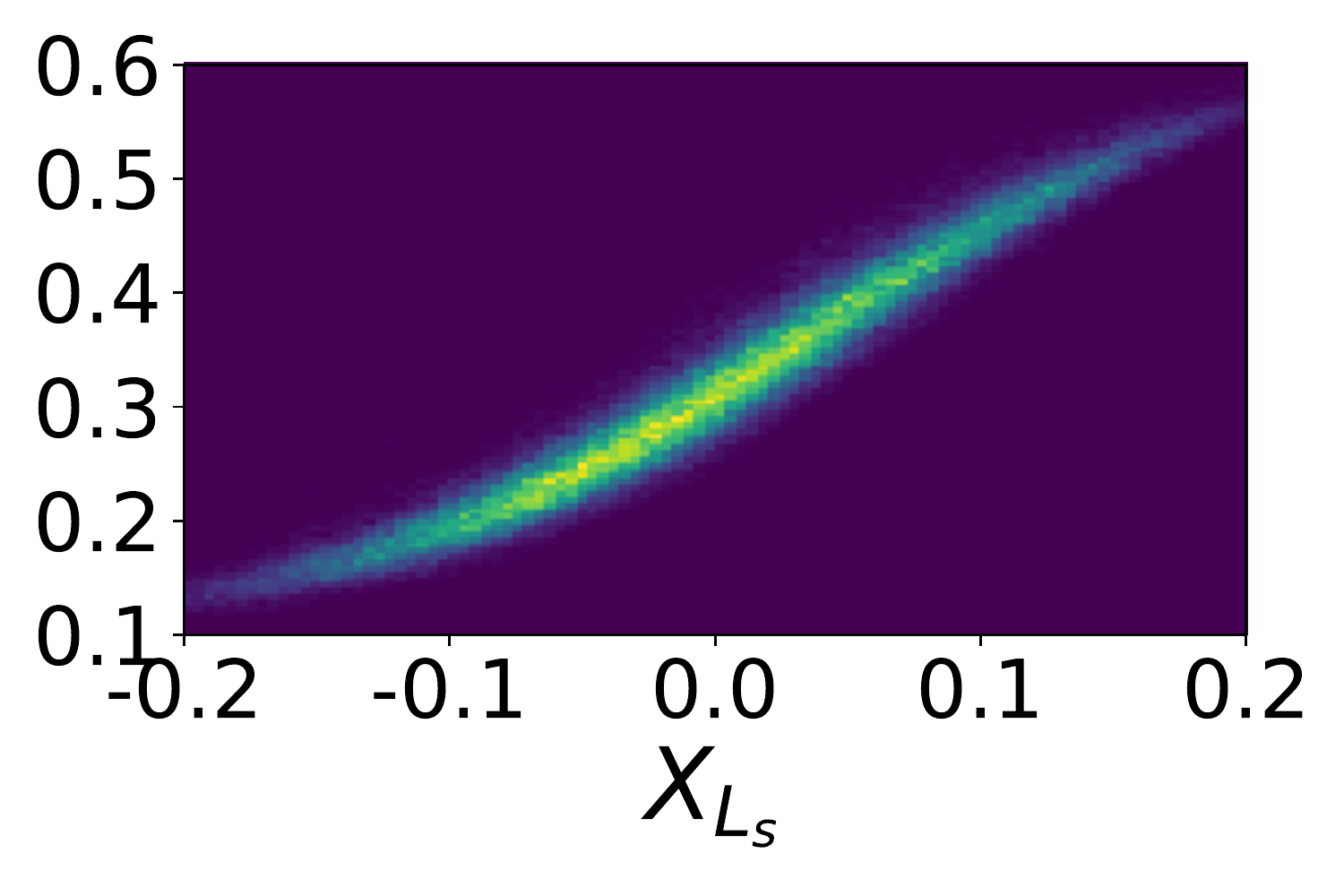,
height=2.2cm,width=2.8cm,angle=0,clip=true,trim = 0.0cm 0.5cm 0.4cm 0.0cm}  \\
\psfig{figure=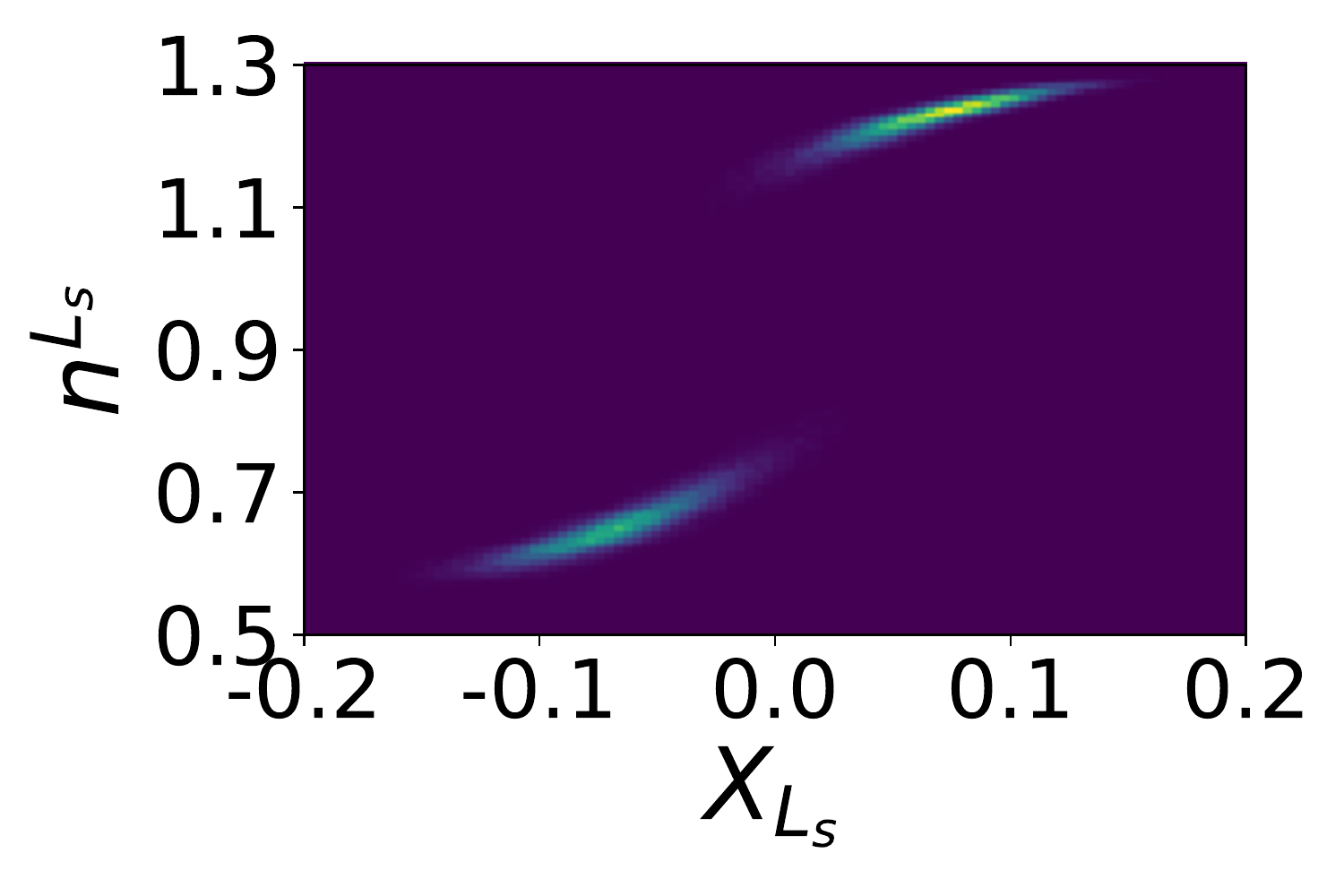,
height=2.2cm,width=2.8cm,angle=0,clip=true,trim = 0.0cm 0.5cm 0.4cm 0.0cm}
\psfig{figure=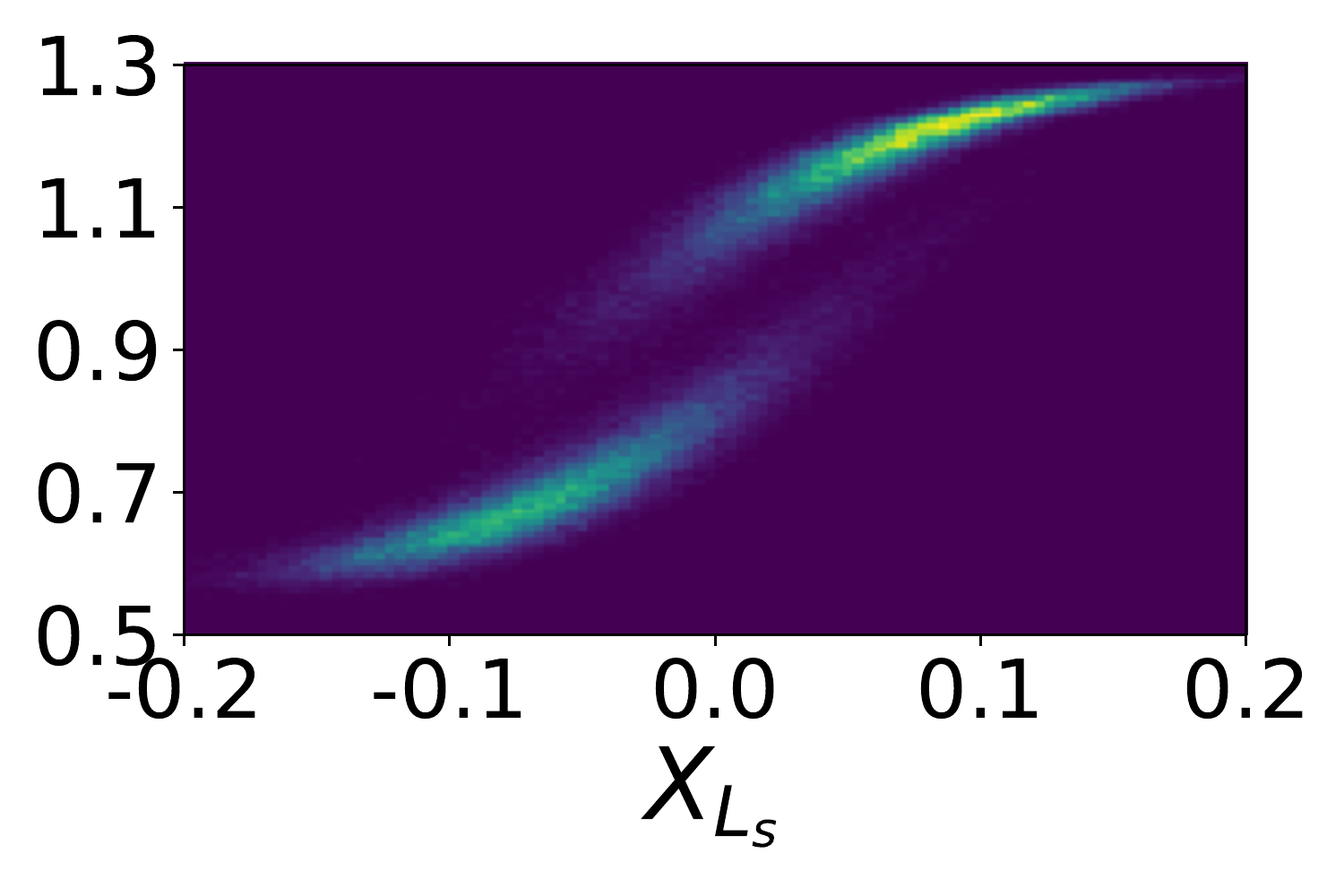,
height=2.2cm,width=2.8cm,angle=0,clip=true,trim = 0.0cm 0.5cm 0.4cm 0.0cm}  
\psfig{figure=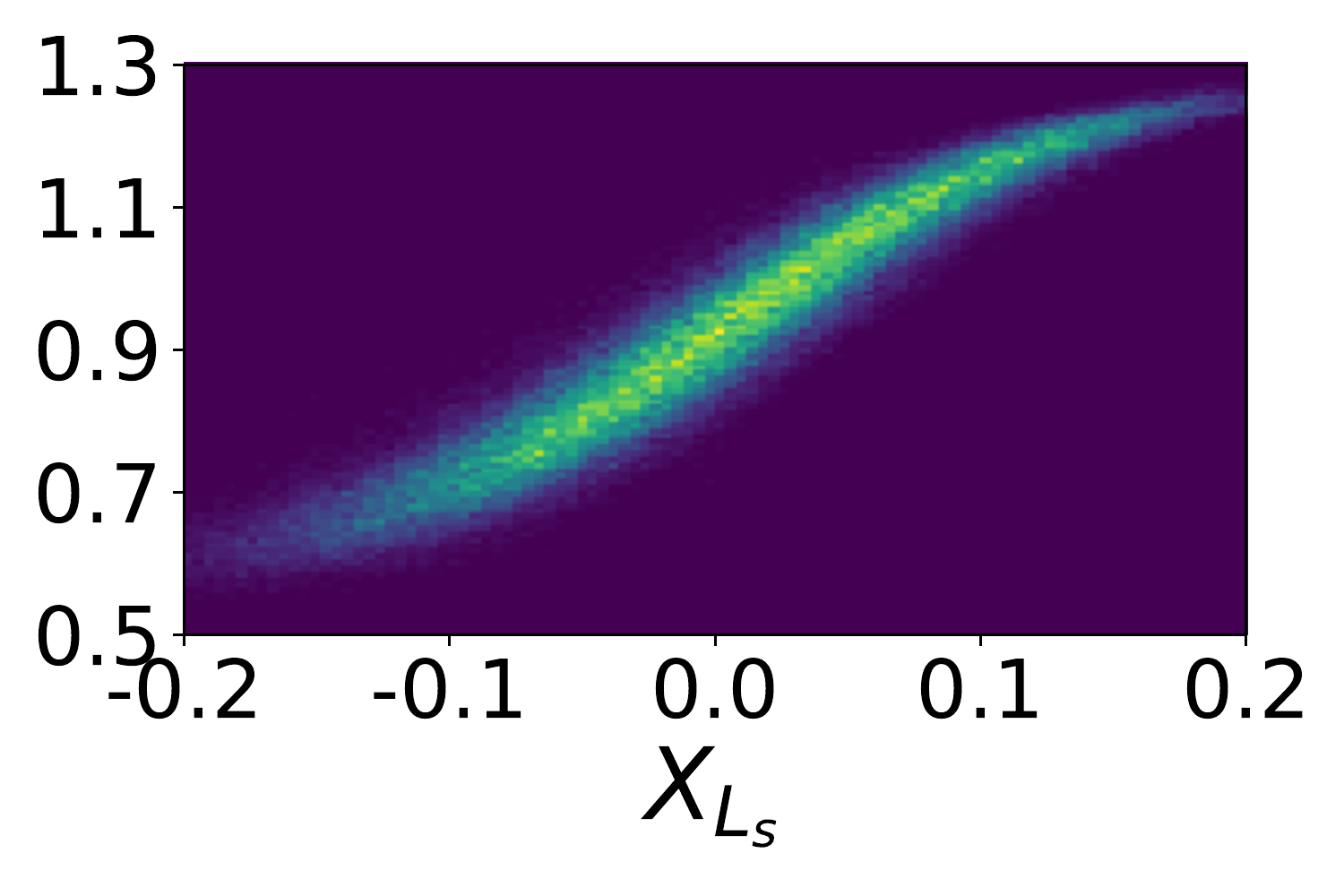,
height=2.2cm,width=2.8cm,angle=0,clip=true,trim = 0.0cm 0.5cm 0.4cm 0.0cm}   \\
\psfig{figure=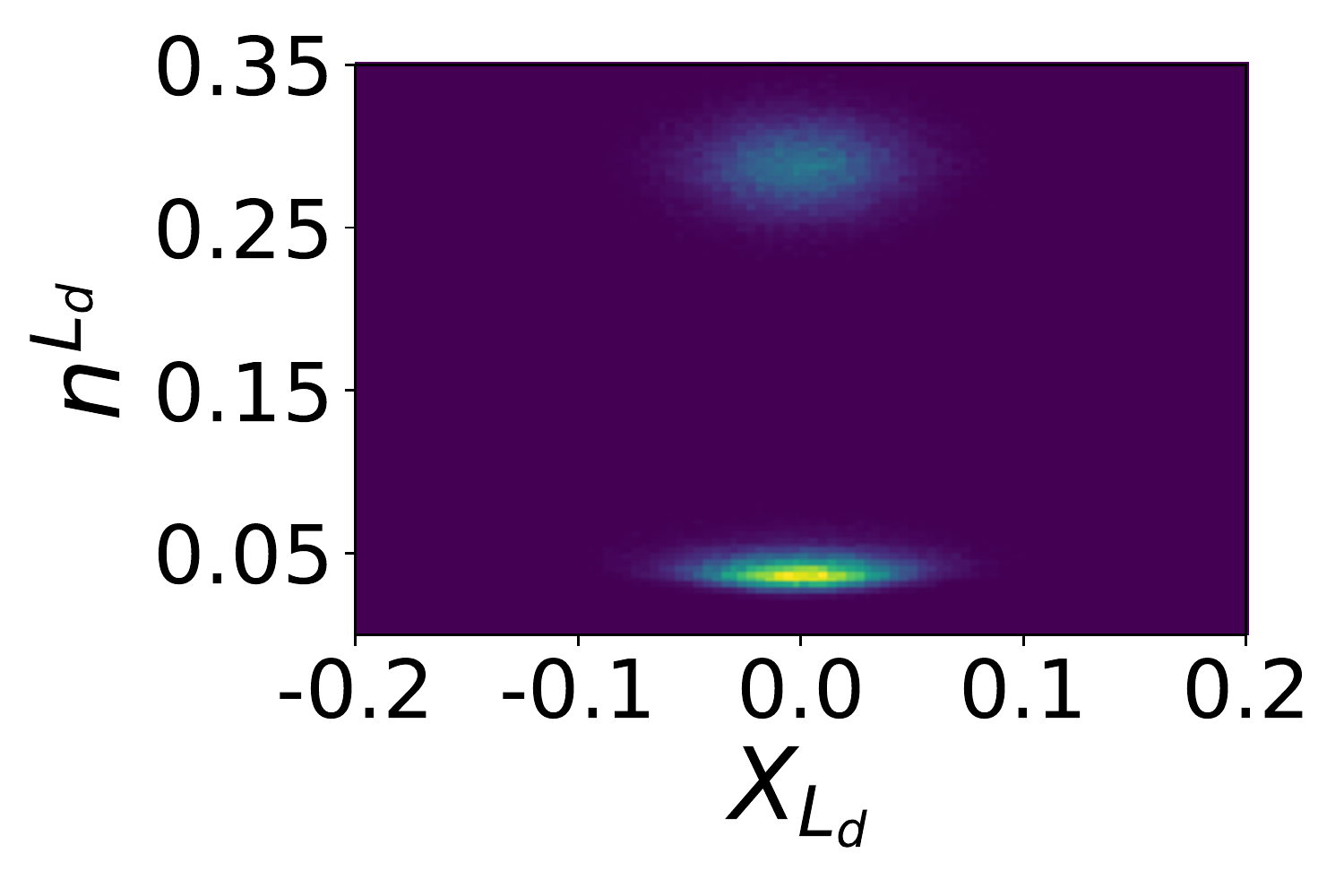,
height=2.2cm,width=2.8cm,angle=0,clip=true,trim = 0.0cm 0.5cm 0.4cm 0.0cm}
\psfig{figure=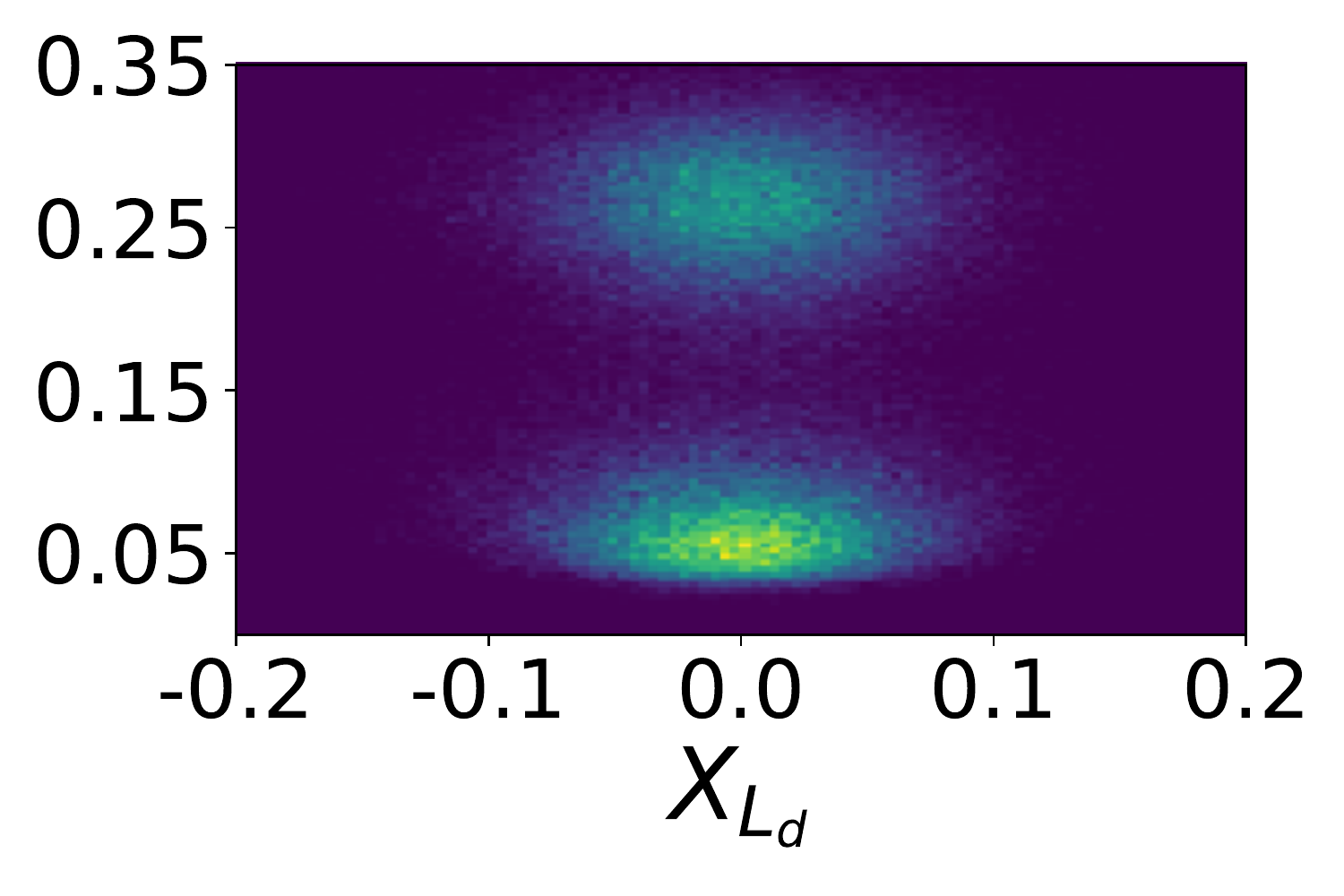,
height=2.2cm,width=2.8cm,angle=0,clip=true,trim = 0.0cm 0.5cm 0.4cm 0.0cm} 
\psfig{figure=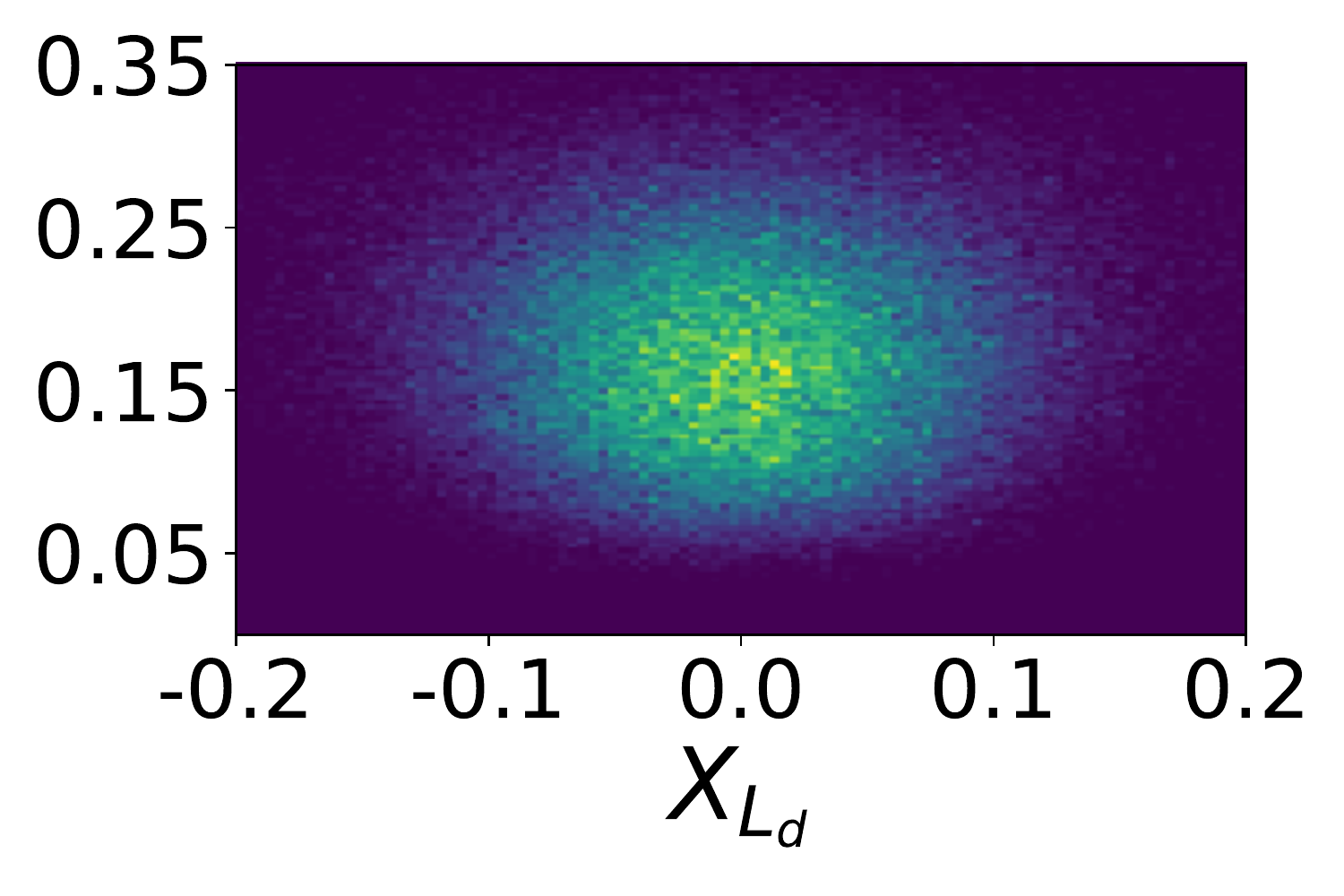,
height=2.2cm,width=2.8cm,angle=0,clip=true,trim = 0.0cm 0.5cm 0.4cm 0.0cm}  
\caption{(Color online) Regression analysis of various local densities along the Monte Carlo step indicates the heating effect of the bond-disproportionation at half-filling (from left to right: $\beta=25, 10, 5$ eV$^{-1}$). Heating effect of the spatiotemporal distribution of four molecular orbitals' occupancies as functions of their corresponding phonon modes at half-filling.}
\label{Fig:Teff1}
\end{figure}

We first concentrate on the effects of heating at half-filling. Fig.~\ref{Fig:Teff1} presents a histogram of the spatiotemporal distribution of different local densities across 1000 independent MC samples. Here, we correlate the orbital occupations of the Bi $6s$ ($n^s$, top row) and ligand oxygen orbitals with $s$ and $d$ symmetry ($n^{L_s}$, middle row and $n^{L_d}$, bottom row, respectively), with the $X_s$ and $X_d$ displacements. Note that our convention in Eq.~\eqref{Xs} implies that a compressed (expanded) oxygen plaquette correspond to a positive (negative) value of $X_{L_s}$. 

The leftmost column of Fig.~\ref{Fig:Teff1} shows typical distributions at low temperature ($\beta=25.0$ eV$^{-1}$). The correlations between $X_{L_s}$ and $n^s$ (top row) and $n^{L_s}$ (middle row) make it clear that the compressed plaquettes host more holes while the expanded plaquettes are hole depleted. 
At the same time, we observe far fewer holes occupying the $n^{L_d}$ (lower row) and $n^{L_{x,y}}$ (not shown), molecular orbitals. Moreover, we find no  correlation between occupations of the $d$, $x$ and $y$ molecular orbitals and the sign of the corresponding phonon displacement, as shown in the bottom row of Fig.~\ref{Fig:Teff1} for the case of the $d$ orbital.

One issue being debated in the context of the charge- and bond-disproportionation scenarios is their relationship to the charge density modulation appearing on the Bi atoms. Our analysis indicates that the occupations of the  Bi orbitals in the expanded and contracted plaquettes are $\sim 0.15$ and $\sim 0.45$, respectively, corresponding to a charge transfer of about $0.2$ holes/Bi in the bond-disproportioned state.  This value is comparable to the $0.1$~holes/Bi obtained using DQMC for a related model~\cite{Steve2020}.

The middle and right columns of Fig.~\ref{Fig:Teff1} shows that the distributions broaden as the temperature increases, reflecting the larger thermal motion of the ions. For $\beta = 10$~eV$^{-1}$, we still observe a bimodal structure in the distributions, albeit with a reduced distance between the two clouds of data points. This behavior indicates that the bond-disproportion correlations are significant but reduced in strength at this temperature. 
Upon increasing the temperature further, the two modes merge into a single distribution as illustrated in the last column, which corresponds to a relatively high temperature of $\beta = 5$ eV$^{-1}$. By scanning the intermediate temperature values, we estimate that the merger occurs around $T\sim0.125$ eV. 
All of these observations are consistent with Fig.~\ref{sublatT}. 

The behavior of the distributions shown in the right hand panel provides strong indications of the persistence of the polarons at high temperature. For example, although the bimodal distributions have merged into a single mode at these temperatures, we observe persistent  correlations between the $n^s$ and $n^{L_s}$ occupations and the sign of the $X_{L_s}$ displacements. Conversely, we do not observe any clear correlation between the occupations of the $n^s$ orbitals and the  $X_{L_d}$ displacements, between the $n_d$ orbitals and the $X_{L_d}$ displacements, or other combinations (not shown). These observations provide compelling evidence that the holes with particular orbital symmetries are strongly attracted to the compressive distortions of the oxygen atoms, consistent with a polaronic view of the metallic phase.  

\subsection{Fingerprints of a bipolaron liquid at high hole concentrations}\label{Sec:Results_Histo2}

\begin{figure}[t] 
\psfig{figure=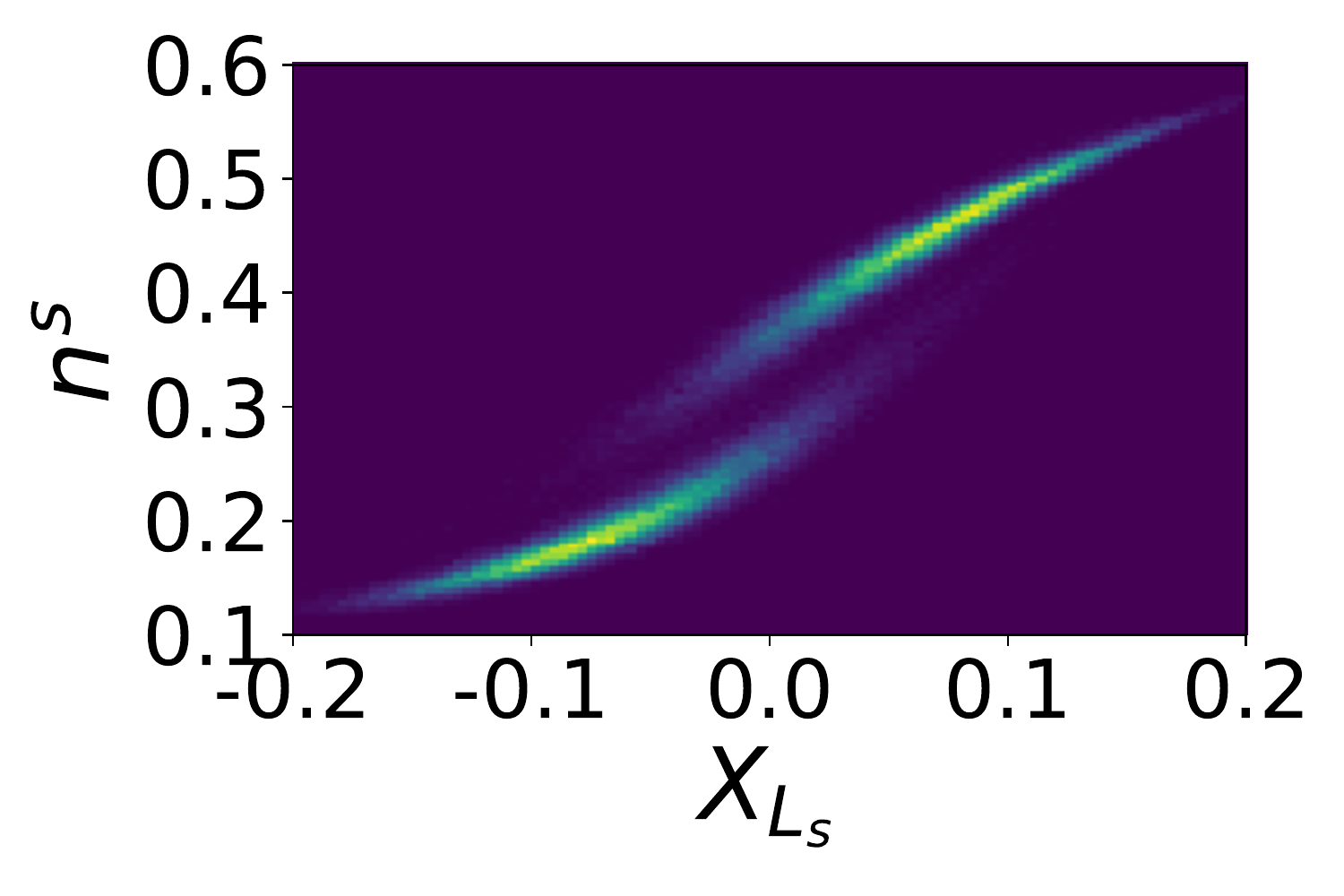,
height=2.2cm,width=2.8cm,angle=0,clip=true,trim = 0.0cm 0.5cm 0.4cm 0.0cm}
\psfig{figure=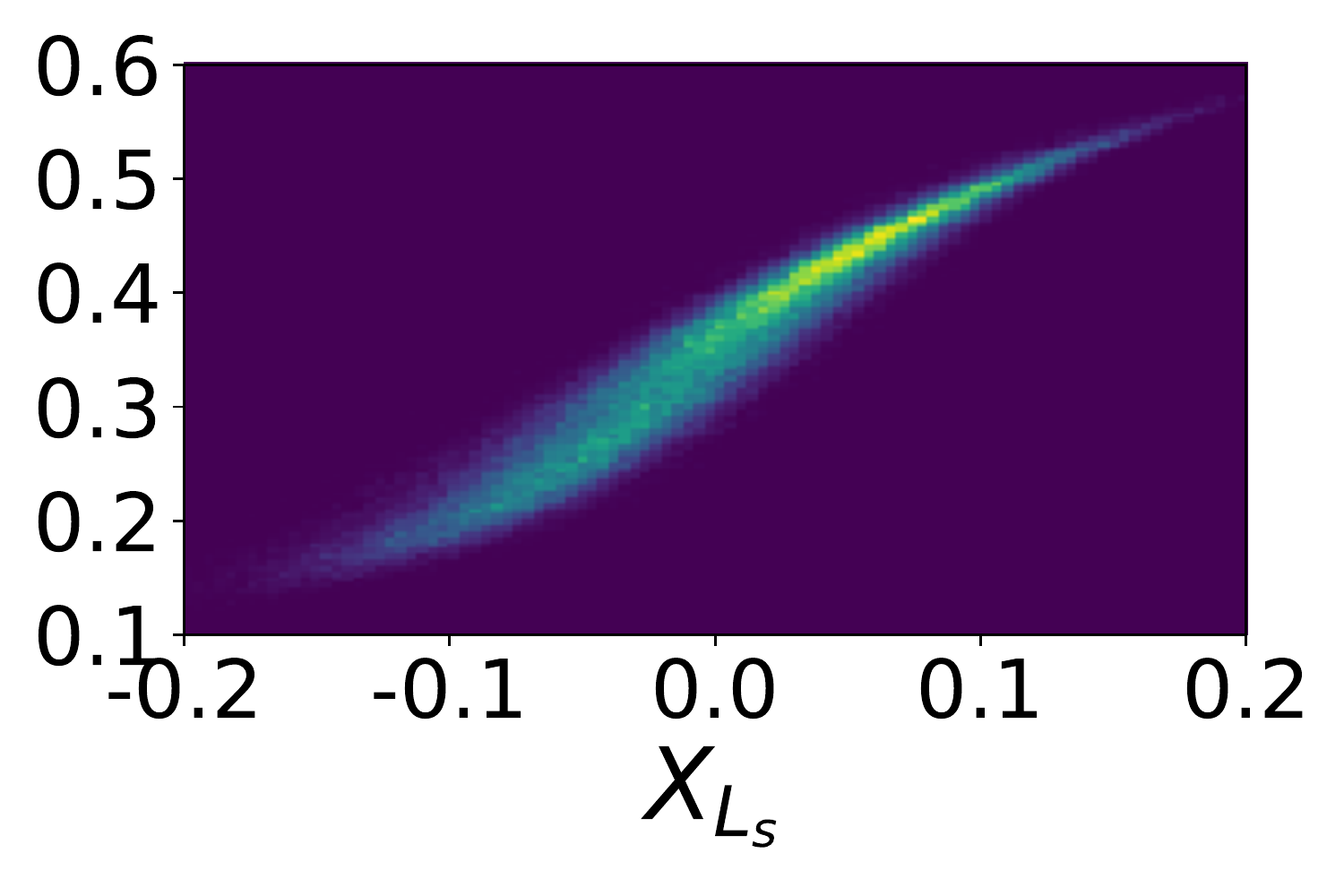,
height=2.2cm,width=2.8cm,angle=0,clip=true,trim = 0.0cm 0.5cm 0.4cm 0.0cm} 
\psfig{figure=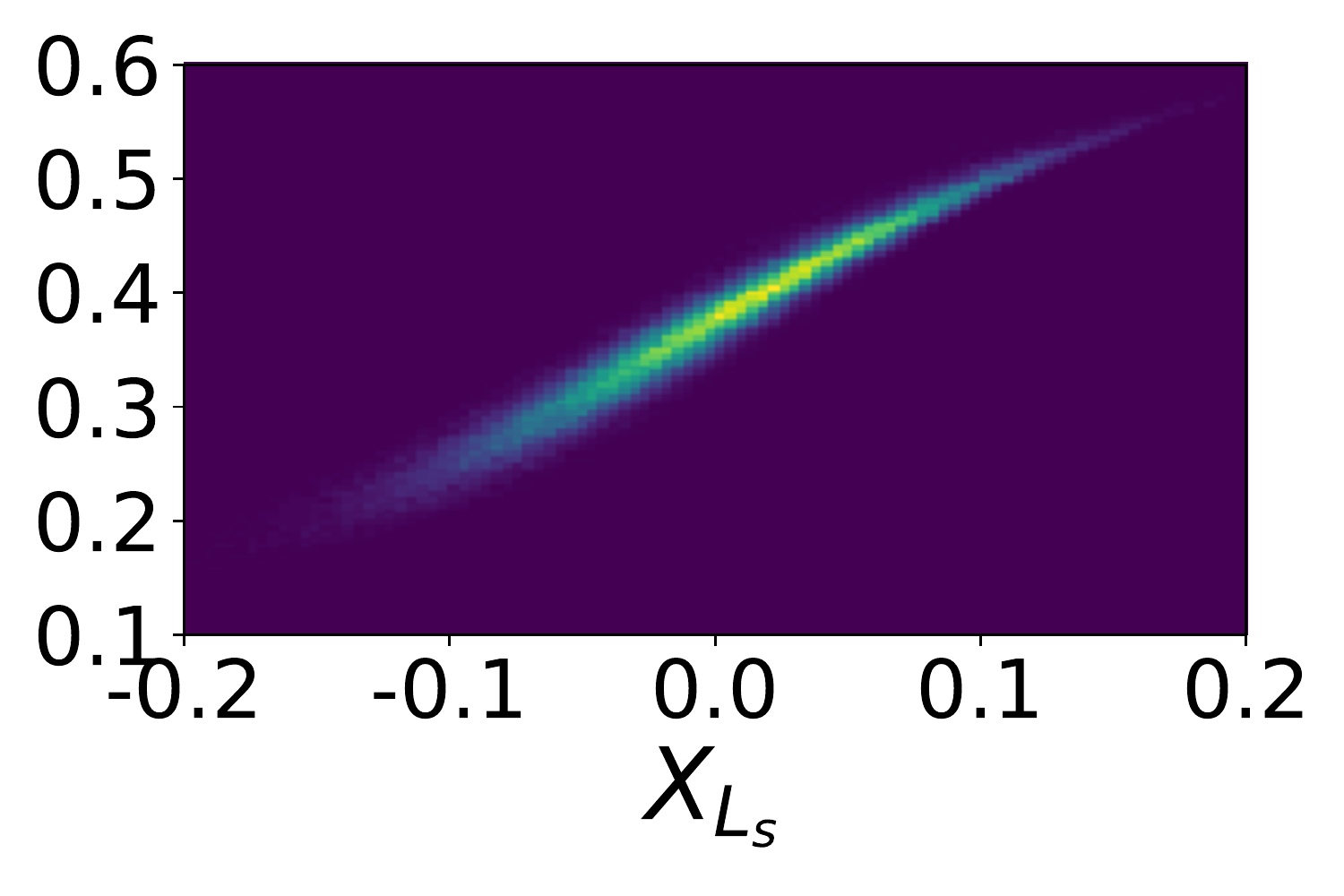,
height=2.2cm,width=2.8cm,angle=0,clip=true,trim = 0.0cm 0.5cm 0.4cm 0.0cm}  \\
\psfig{figure=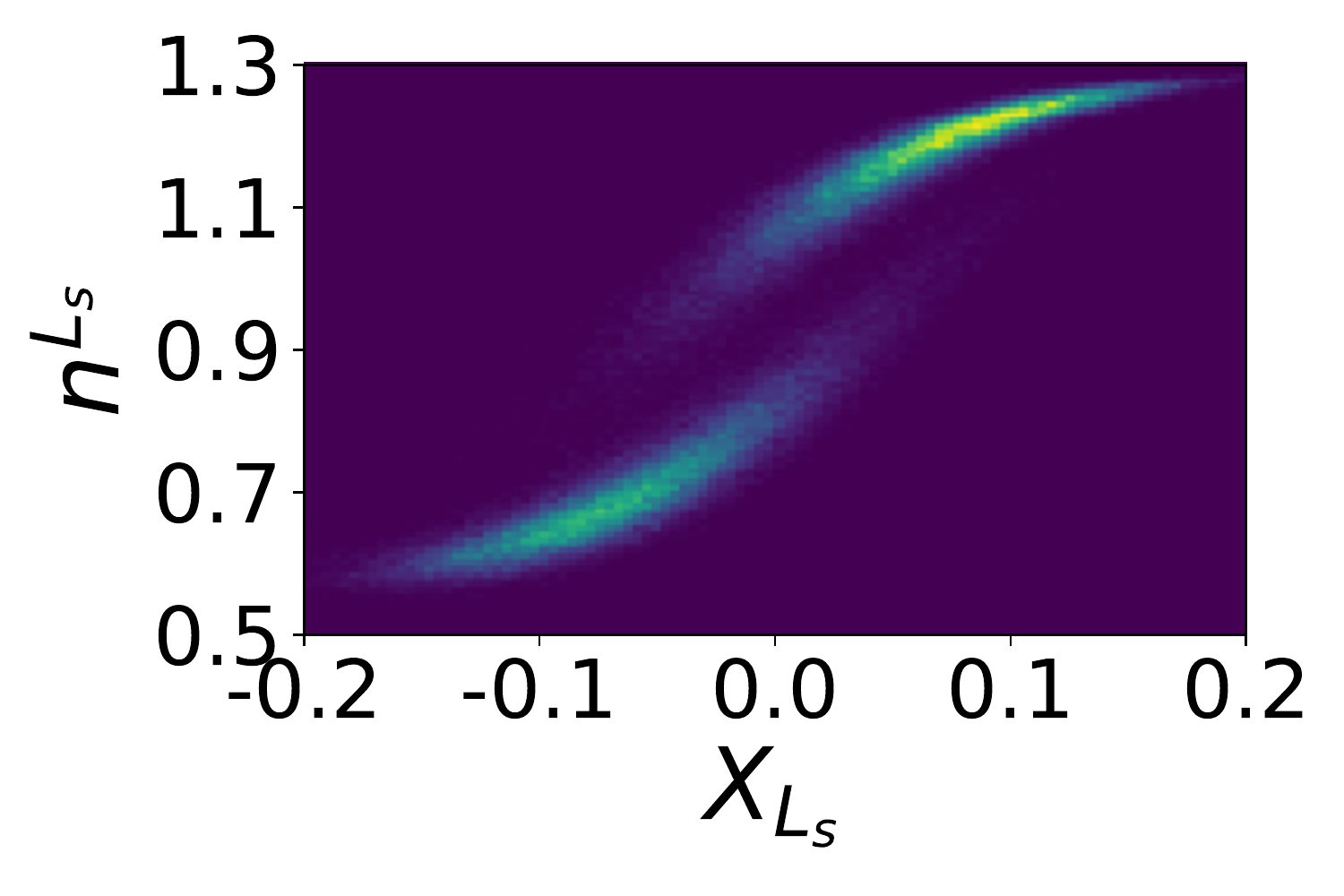,
height=2.2cm,width=2.8cm,angle=0,clip=true,trim = 0.0cm 0.5cm 0.4cm 0.0cm}  
\psfig{figure=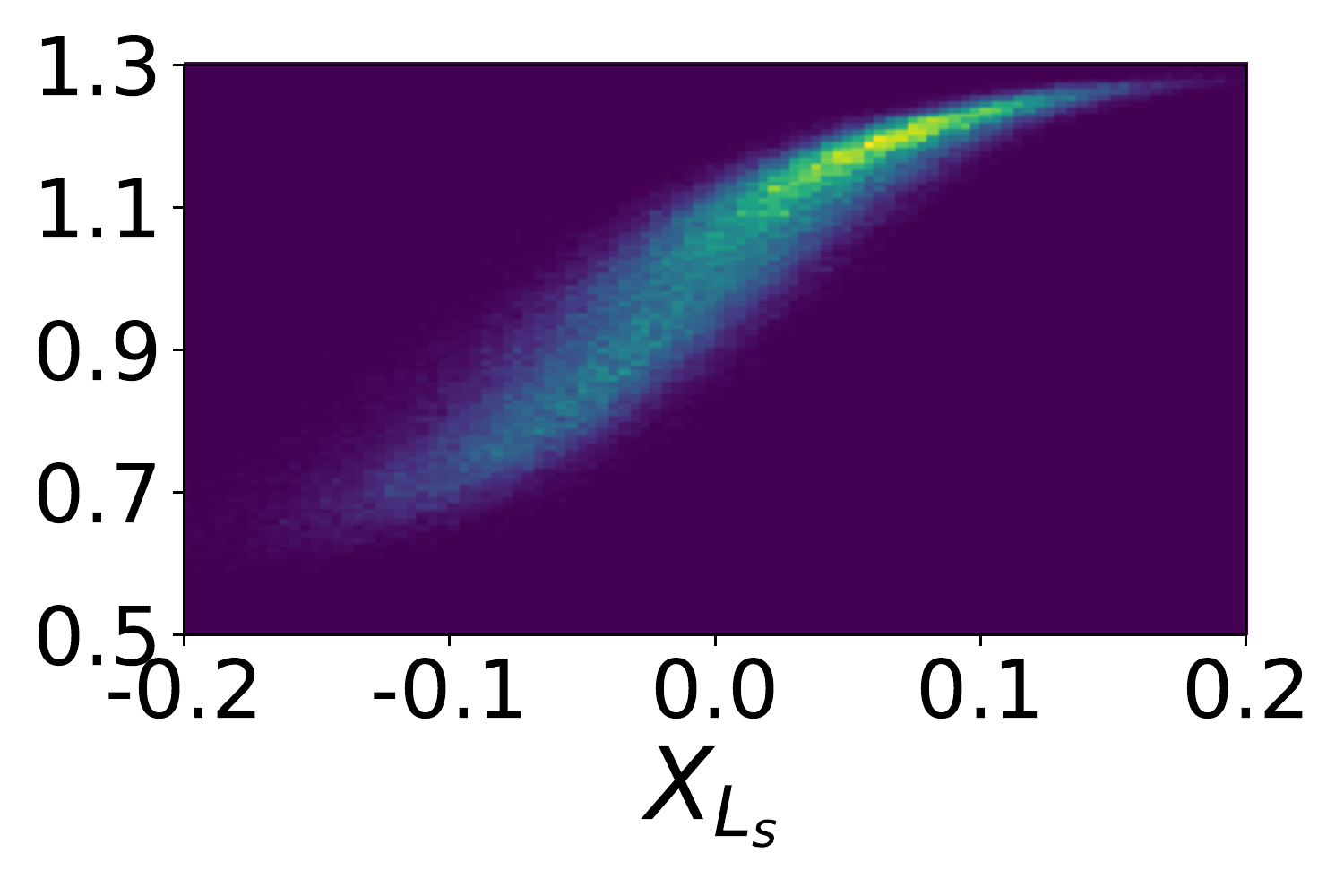,
height=2.2cm,width=2.8cm,angle=0,clip=true,trim = 0.0cm 0.5cm 0.4cm 0.0cm}  
\psfig{figure=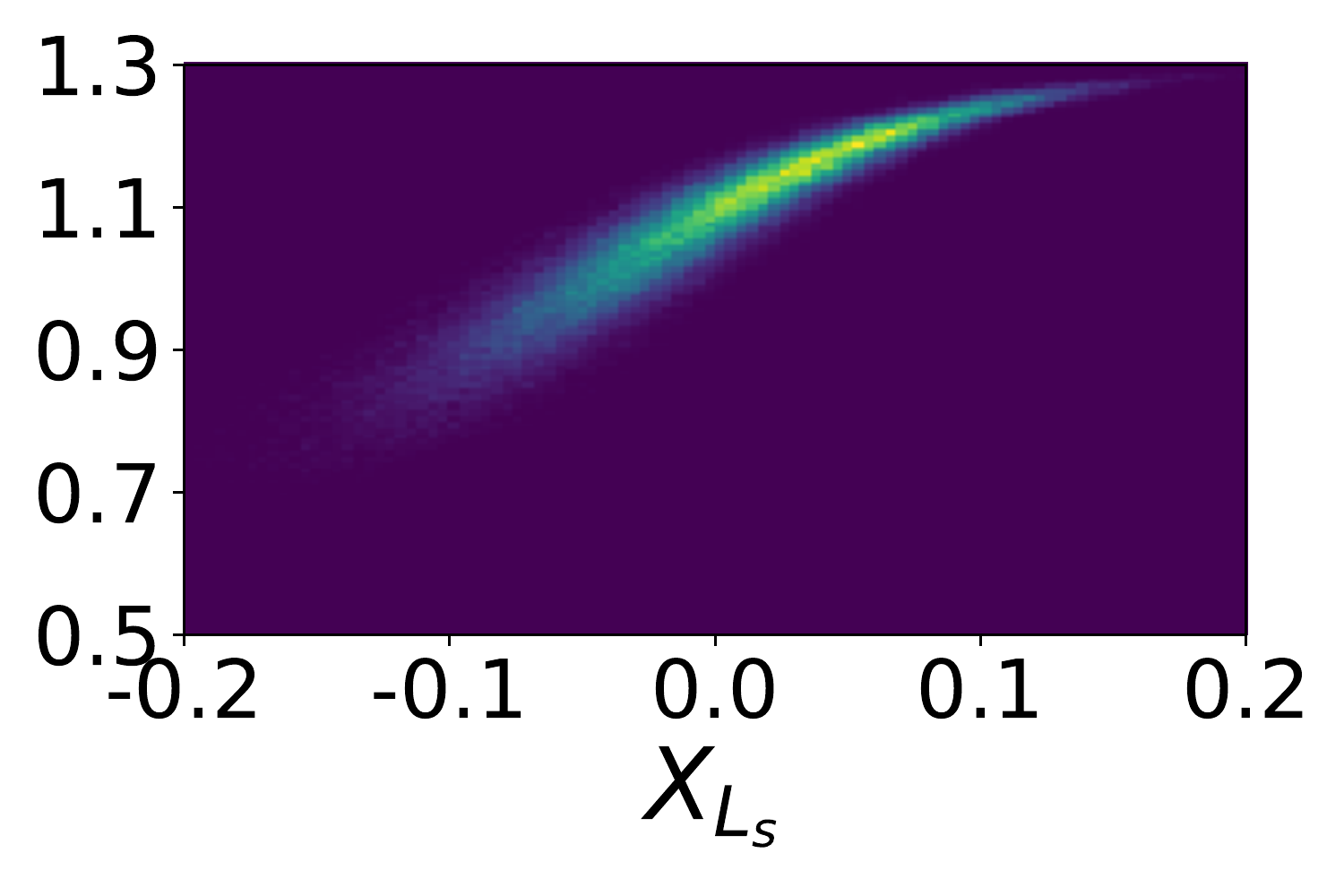,
height=2.2cm,width=2.8cm,angle=0,clip=true,trim = 0.0cm 0.5cm 0.4cm 0.0cm}    \\
\psfig{figure=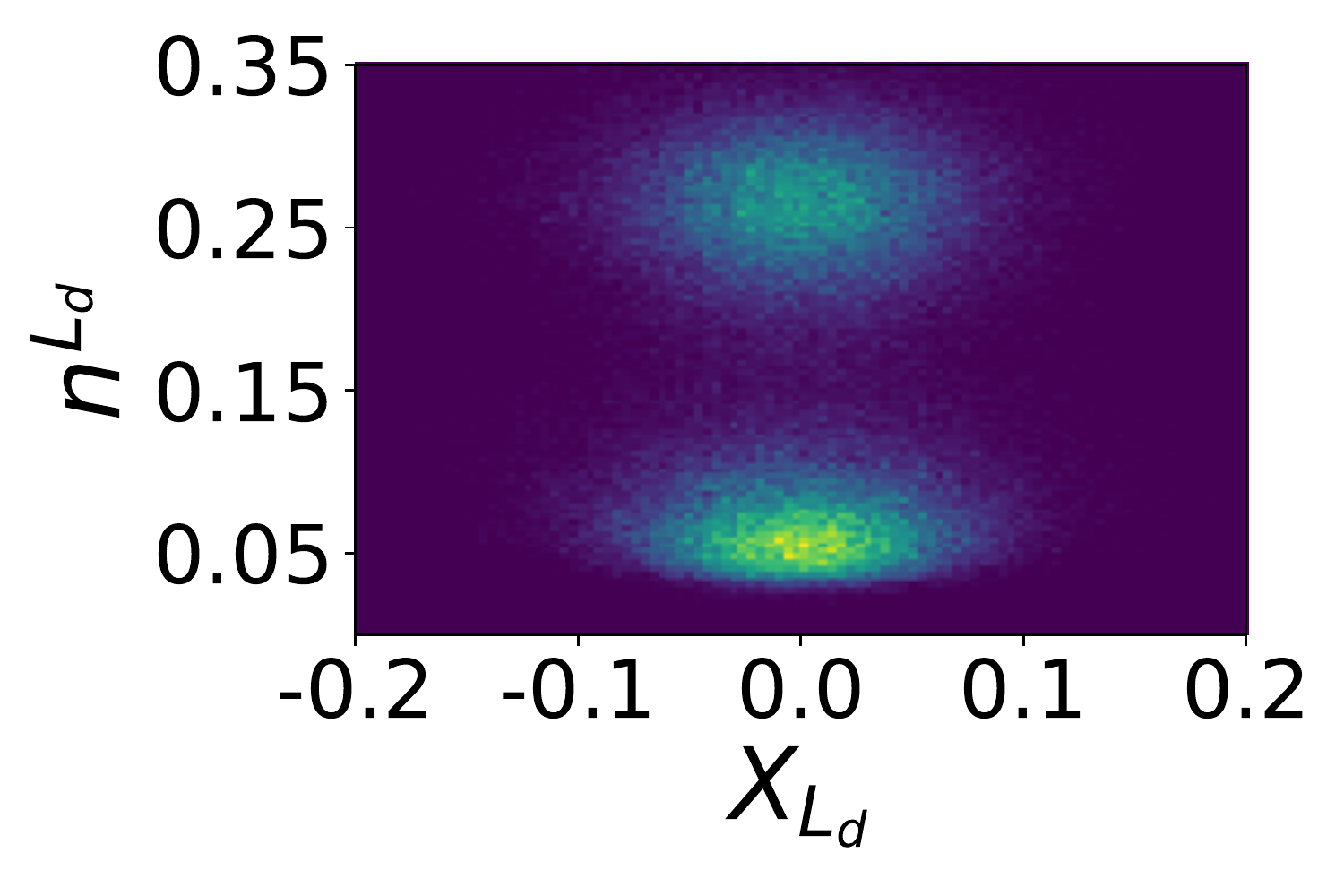,
height=2.2cm,width=2.8cm,angle=0,clip=true,trim = 0.0cm 0.5cm 0.4cm 0.0cm}  
\psfig{figure=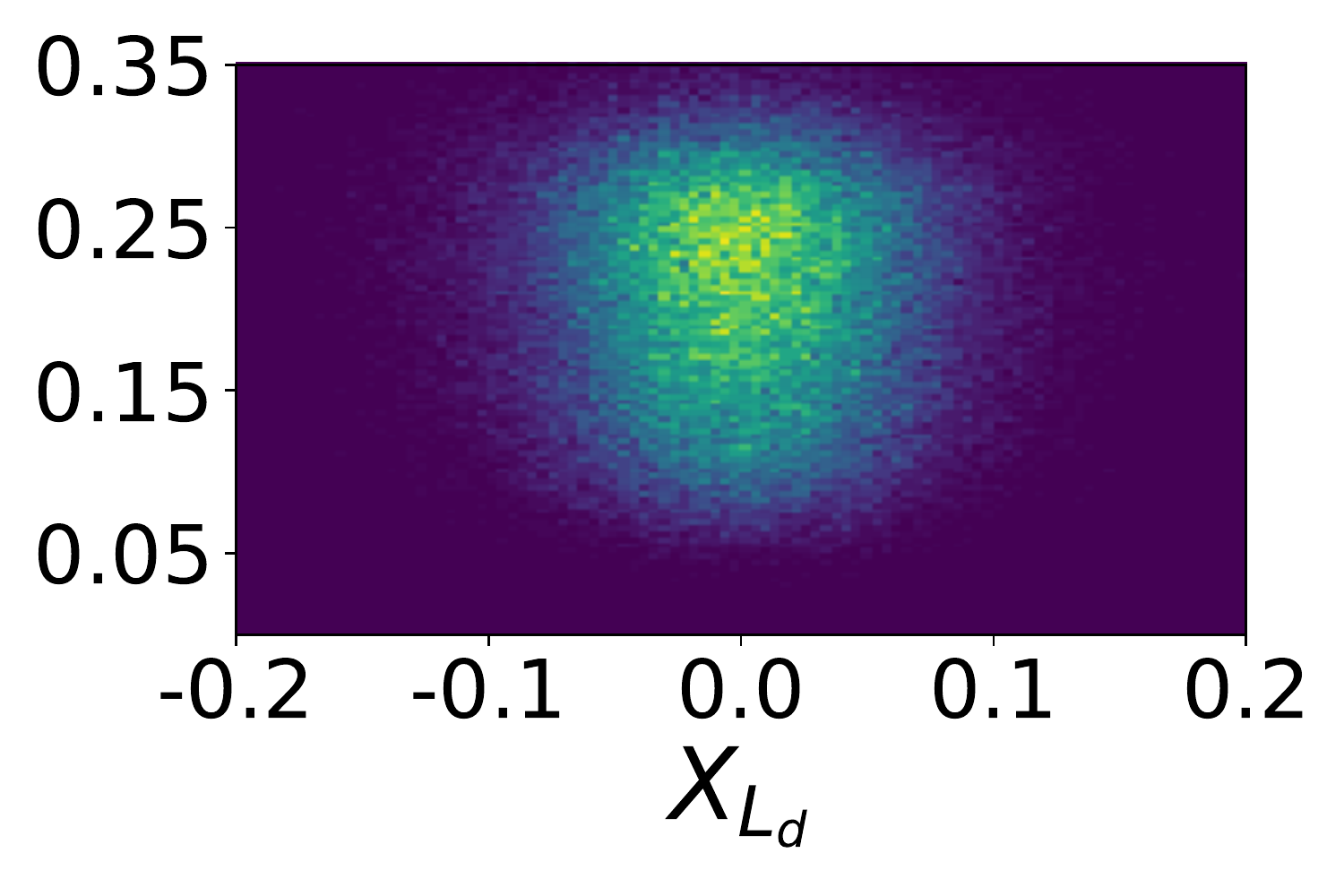,
height=2.2cm,width=2.8cm,angle=0,clip=true,trim = 0.0cm 0.5cm 0.4cm 0.0cm}  
\psfig{figure=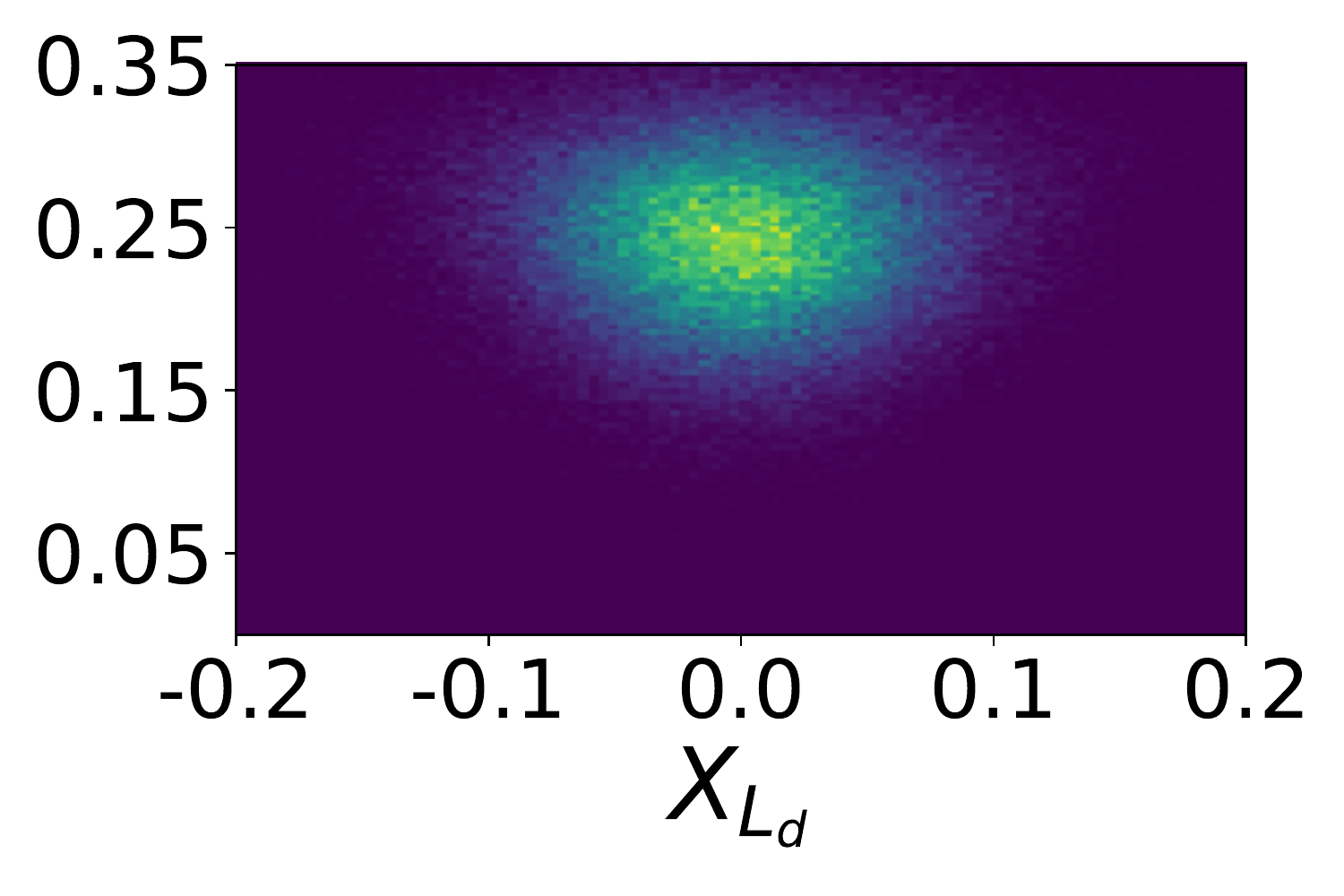,
height=2.2cm,width=2.8cm,angle=0,clip=true,trim = 0.0cm 0.5cm 0.4cm 0.0cm}     
\caption{(Color online) Doping effect at $\beta=20.0$ (from left to right: 0\%, 10\%, 20\% doped holes approximately): regression analysis of various densities along the Monte Carlo step for phonon frequency $\Omega=30$ meV for $N=10\times 10, t_{sp} = 2.08$, $t_{pp} = 0.056$, $\epsilon_s = 6.42$, and $\epsilon_p = 2.42$.}
\label{Fig:dope1}
\end{figure}

We now examine the doping-driven transition from the bond-disproportionate state to the uniform metallic state using a similar spatiotemporal distribution analysis. To this end, Fig.~\ref{Fig:dope1} plots the evolution of the same distributions shown in Fig.~\ref{Fig:Teff1}, this time as a function of hole concentration and at a fixed temperature $\beta=20.0$ eV$^{-1}$. (We observe qualitatively similar results at a fixed $\beta=10.0$ eV$^{-1}$ and will, therefore, focus exclusively on the lower temperature case.)

The leftmost column of Fig.~\ref{Fig:dope1}  shows the typical distribution at half-filling, similar to the leftmost column of Fig.~\ref{Fig:Teff1}. Because $\beta=20.0$ is lower than $T_c=0.125$ eV in Fig.~\ref{sublatT}, its two-cloud feature is again consistent with the bond-disproportionation. 

The middle and right columns illustrate the doping induced melting,  which is reminiscent of that induced by heating, shown in Fig.~\ref{Fig:Teff1}. The major difference from the melting by heating is the global shift upwards of the distributions, due to the doping induced increase of $n^s$ and $n^{L_s}$.
Apart from that, the transition from the two-clouds to the one-cloud distribution is rather similar in both cases; all this is consistent with the data shown in Fig.~\ref{sublatT}. Like in Fig.~\ref{Fig:Teff1}, the persistence of correlations between hole occupations in the bonding orbital and the compression of the corresponding plaquette, suggests that the metallic state induced by doping is also consistent with a melted liquid of (bi)polarons. The associated short-range spatial correlations are dynamically averaged over time, so that long-time averages look like those of a homogeneous system, as shown in  Fig.~\ref{sublatT}.

\section{Summary and Conclusions}\label{Sec:Conclusion}
We have presented a numerical study of a two-dimensional three-orbital model with SSH-like $e$-ph interactions, which was solved using a combined exact diagonalization and classical Monte Carlo method (ED + MC). Having bismuthates in mind, we studied the model for concentrations near one hole/cation site, and in the negative charge transfer regime, where the holes preferably occupy the oxygen sublattice. Our key results include: 1) the observation of a bond-disproportionated insulating state at half-filling, which can be viewed as the frozen bipolaronic crystal state; 2) the observation of insulator-to-metal transitions with either increasing temperature or hole doping; and 3) the observation of local bipolaron formation in the resulting metallic phases.  

While many of our conclusions agree with a recent DQMC study of a similar model~\cite{Steve2020}, our work should be seen as complementary to this study. For example, by using classical treatment of the lattice degrees of freedom, we accessed much larger system sizes and smaller phonon energies. This aspect places our model in the adiabatic regime, which is more relevant to the real materials, while Ref.~\onlinecite{Steve2020} considered a set of anti-adiabatic phonon parameters. As such, our results enlarge the parameter space considerably and provide a more complete view of the physics of the model. In some instances, we were even able to perform meaningful extrapolations to the thermodynamic limit. Moreover, our spatiotemporal analysis allowed us to examine the correlations between various local electronic quantities and the lattice degrees of freedom, which in turn allowed us to identify the similarities 
in the heating- and doping-induced melting of the bond-disproportionated state. 
Contrary to the previous DQMC study, however, our analysis of local charge density on Bi ions observed a more significant charge modulation between the Bi atoms in alternating compressed and expanded octahedra, which is even higher than the values inferred experimentally~\cite{Hair1973,Orchard1977,Wertheim1982,Plumb2016}. This inconsistency may be related to the classical treatment of the lattice vibrations or the fact that we have neglected a small local Hubbard repulsion on the Bi sites. 

\section{Acknowledgements}
M.~J., M.~B., and G.~A.~S. acknowledge support from the Stewart Blusson Quantum Matter Institute, Natural Sciences and Engineering Research Council (NSERC) for Canada, and Canada First Research Excellence Fund (CFREF). S.~J. acknowledges support from the Scientific Discovery through Advanced Computing (SciDAC) program funded by the U.S. Department of Energy, Office of Science, Advanced Scientific Computing Research and Basic Energy Sciences, Division of Materials Sciences and Engineering.


\end{document}